\documentclass[12pt,english,onecolumn, draftcls]{IEEEtran}

\usepackage[T1]{fontenc}
\usepackage[latin1]{inputenc}
\usepackage{geometry}
\usepackage{amsmath,amssymb,amsthm,mathrsfs,amsfonts,dsfont}
\usepackage{graphicx}
\usepackage{epstopdf}
\usepackage{cite}
\usepackage{color}
\usepackage{mathtools}
\usepackage{cases}
\usepackage{stackengine}
\usepackage{accents}
\usepackage{caption}
\usepackage{subcaption}
\usepackage{float}
\usepackage{verbatim}
\usepackage{multirow}
\usepackage{cancel}
\usepackage[shortlabels]{enumitem}
\DeclareMathOperator*{\argmin}{argmin}
\newcommand\blfootnote[1]{%
\begingroup
\renewcommand\thefootnote{}\footnote{#1}%
\addtocounter{footnote}{-1}%
\endgroup}

\newtheorem{rem}{Remark}

\renewcommand\appendix{\par
\setcounter{section}{0}
\setcounter{subsection}{0}
\setcounter{figure}{0}
\setcounter{table}{0}
\renewcommand\thesection{ \Alph{section}}
\renewcommand\thefigure{\Alph{section}\arabic{figure}}
\renewcommand\thetable{\Alph{section}\arabic{table}}
}

\definecolor{Jakob}{rgb}{1.0, 0.75, 0.0}	
\definecolor{Bruno}{rgb}{1.0, 0.13, 0.32}
\definecolor{come}{rgb}{0.82, 0.62, 0.91}

\usepackage{babel}

\allowdisplaybreaks

\begin{document}

\title{Learning to Communicate and Energize: Modulation, Coding and Multiple Access Designs for Wireless Information-Power Transmission}

\author{Morteza Varasteh, Jakob Hoydis and Bruno Clerckx\\
\thanks{M. Varasteh is with School of Computer Science and Electronic Engineering, University of Essex, UK (email: m.varasteh@essex.ac.uk). J. Hoydis is with Nokia Bell Labs, Nozay, France (email: jakob.hoydis@nokia-bell-labs.com). B. Clerckx is with the EEE department at Imperial College London, London SW7 2AZ, UK (email: b.clerckx@imperial.ac.uk).}\thanks{This work has been partially supported by the EPSRC of the UK under grant EP/P003885/1.
}}

%%%%%%%%%%%%%%%%%%%
\maketitle
\vspace{-13mm}
\begin{abstract}

The explosion of the number of low-power devices in the next decades calls for a re-thinking of wireless network design, namely, unifying wireless transmission of information and power so as to make the best use of the RF spectrum, radiation, and infrastructure for the dual purpose of communicating and energizing. This paper provides a novel learning-based approach towards such wireless network design. To that end, a parametric model of a practical energy harvester, accounting for various sources of nonlinearities, is proposed using a nonlinear regression algorithm applied over collected real data. Relying on the proposed model, the learning problem of modulation design for \textit{Simultaneous Wireless Information-Power Transmission} (SWIPT) over a point-to-point link is studied. Joint optimization of the transmitter and the receiver is implemented using \textit{Neural Network} (NN)-based autoencoders. The results reveal that by increasing the receiver power demand, the baseband transmit modulation constellation converges to an On-Off keying signalling. Utilizing the observations obtained via learning, an algorithmic SWIPT modulation design is proposed. It is observed via numerical results that the performance loss of the proposed modulations are negligible compared to the ones obtained from learning. Extension of the studied problem to learning modulation design for multi-user SWIPT scenarios and coded modulation design for point-to-point SWIPT are considered. The major conclusion of this work is to utilize learning-based results to design non learning-based algorithms, which perform as well. In particular, inspired by the results obtained via learning, an algorithmic approach for coded modulation design is proposed, which performs very close to its learning counterparts, and is significantly superior due to its high real-time adaptability to new system design parameters\blfootnote{This work has been partially presented in the IEEE International Conference on Acoustics, Speech and Signal Processing (ICASSP) 2019 \cite{Varasteh_Piovano_Clerckx}, and IEEE International Workshop on Signal Processing Advances in Wireless Communications (SPAWC) 2019 \cite{Varasteh_Hoydis_Clerckx}.}.
\end{abstract}

\vspace{-1mm}
\section{Introduction}\label{Sec_Intro}

\textit{Radio Frequency} (RF) signals are capable of bearing information as well as power. The transferred power can be utilized for energizing low-power devices, such as wireless sensors and \textit{Internet-of-Things} (IoT) devices. This along with the growth of low-power devices has created a significant attention towards the study of \textit{Simultaneous Wireless Information-Power Transmission} (SWIPT) systems \cite{Varshney_2008, Zhang_Keong_2013, Clerckx_Zhang_Schober_Wing_Kim_Vincent}. In order to design efficient SWIPT architectures, it is crucial to model the \textit{Energy Harvester} (EH) accurately. The EH consists of a rectenna, which is composed of an antenna followed by a rectifier. The rectifier is used to convert the RF power into DC power in order to charge devices. Although most of the results in the literature adopt a linear characteristic function for the rectifier, in practice, due to the presence of a diode in the rectifier, the output of the EH is a nonlinear function of its input \cite{Clerckx_Bayguzina_2016, Boaventura_Collado_Carvalho}.

Due to the nonlinearity of the diode characteristic function, the RF-to-DC conversion efficiency of the EH is highly dependent on the power as well as the shape of the input waveform \cite{Clerckx_Bayguzina_2016,Boaventura_Collado_Carvalho,Clerckx_Bayguzina_2017}. Observations based on experimental results reveal that signals with high \textit{Peak-to-Average Power Ratio} (PAPR) result in high delivered DC power compared to other signals \cite{Boaventura_Collado_Carvalho}. Motivated by this observation, in \cite{Clerckx_Bayguzina_2016}, an analytical model for the rectenna is introduced and a joint optimization over the phase and amplitude of a deterministic multi-sine signal is studied. It is concluded that unlike a linear EH model that favours single-carrier transmissions, a nonlinear model favours multi-carrier transmissions.

In \textit{Wireless Power Transfer} (WPT) systems, the goal is to design waveforms that maximize the DC power at the output of the EH, whereas, in SWIPT systems, the goal is to maximize the DC power as well as the information rate, which is commonly referred as maximizing the \textit{Rate-Power} (RP) region. Unlike most of the SWIPT systems with the linear model assumption for an EH, for SWIPT systems with nonlinear EH, there exists a tradeoff between the rate and the delivered power \cite{Clerckx_Zhang_Schober_Wing_Kim_Vincent}. Due to the presence of nonlinear components in an EH, obtaining the exact optimal tradeoff analytically has so far been unsuccessful. However, after making some simplifying assumptions, some interesting results have been derived in \cite{Clerckx_2016,Varasteh_Rassouli_Clerckx_ITW_2017,Morsi_Jamali,Varasteh_Rassouli_Clerckx_arxiv}. In particular, in multi-carrier transmissions, it is shown in \cite{Clerckx_2016} that nonzero mean Gaussian input distributions lead to an enlarged RP region compared to \textit{Circularly Symmetric Complex Gaussian} (CSCG) input distributions. In single carrier transmissions over \textit{Additive White Gaussian Noise} (AWGN) channel, in  \cite{Varasteh_Rassouli_Clerckx_arxiv,Morsi_Jamali}, it is shown that (under nonlinearity assumption for an EH) for circular symmetric inputs, the capacity achieving input distribution is discrete in amplitude with a finite number of mass-points and with a uniformly distributed independent phase. This is in contrast to the linear model assumption of an EH, where there is no tradeoff between the information and power (i.e., from system design perspective the two goals are aligned), and the optimal inputs are Gaussian distributed \cite{Clerckx_Zhang_Schober_Wing_Kim_Vincent}.

While designing SWIPT signals and systems (under nonlinear assumptions for an EH) using analytical tools seems extremely cumbersome, \textit{Deep Learning} (DL)-based methods reveal a promising alternative to tackle the aforementioned problems. In fact, DL-based methods, and particularly, denoising autoencoders have recently shown remarkable results in communications, achieving or even surpassing the performance of state-of-the-art algorithms \cite{OShea_Hoydis_2017,OShea_Karra_Clancy_2016}. The advantage of DL-based methods versus analytical algorithms lies in their ability to extract complex features from training data, and the fact that their model parameters can be trained efficiently on large datasets via backpropagation. The DL-based methods learn the statistical characteristics from a large training dataset, and optimize the algorithm accordingly, without obtaining explicit analytical results. At the same time, the potential of DL has also been capitalized by researchers to design novel and efficient coding and modulation techniques in communications. In particular, the similarities between the autoencoder architecture and the digital communication systems have motivated significant research efforts in the direction of modelling end-to-end communication systems using the autoencoder architecture \cite{OShea_Hoydis_2017,OShea_Karra_Clancy_2016}. Some examples of such designs include decoder design for existing channel codes \cite{Nachmani_etall}, blind channel equalization \cite{Caciularu_Burshtein}, learning physical layer signal representation for SISO \cite{OShea_Hoydis_2017} and MIMO systems \cite{Timothy_OShea_Erpek} and OFDM systems \cite{Felix_Cammerer_Dorner,Ye_Li_Juang}.

In this work, we leverage the learning approach to communicate and energize wirelessly by developing DL-based methods to design modulation, coding, and multiple access for SWIPT networks. In particular, we consider a SWIPT system as a denoising autoencoder structure, where the transmitter(s) and the receiver(s) are considered as multi-layer Deep Neural Networks (DNN). The results are obtained by optimizing the transmitter(s) and receiver(s) jointly over a large training data. Together with the conference papers \cite{Varasteh_Piovano_Clerckx,Varasteh_Hoydis_Clerckx}, this is the first work that investigates the design of SWIPT using machine learning techniques. The contributions of this manuscript are as follow.
\begin{itemize}
\item \textit{Parametric EH model:} A new model for an EH is proposed by applying nonlinear regression algorithm over collected real data corresponding to different input-power/output-power of a practical EH. The proposed parametric model is advantageous compared to the previous models proposed in the literature due to the fact that, besides being simple in structure, it captures (in a unified manner) both challenging limitations of a practical EH, i.e., \textit{saturation effect} (power level above which an EH is saturated) and \textit{threshold effect} (power level below which an EH is off).

\item \textit{Learning point-to-point SWIPT modulations:} Learning modulation design for a point-to-point SWIPT (PP-SWIPT) over a noisy channel is considered using the proposed model for an EH. It is observed that in the low input power range where the \textit{saturation effect} does not occur, namely below $317~\mu$W, some of the modulation symbols move away from zero amplitude as the receiver power demand increases (equivalently, forcing the other symbols converging towards zero amplitude). The number of symbols that move away from zero is dependent on the EH average input power. In the very extreme scenario, where the receiver power demand is at its maximum, some symbols collapse on top of each other at zero amplitude with the others located around a circle in an equidistance manner. This signalling is similar to an On-Off keying signalling and is inline with the result in \cite{Varasteh_Rassouli_Clerckx_arxiv}, where it is shown that under low input average power constraints (the power range where an EH is not saturated) and for power delivery purposes, the optimal capacity achieving input is On-Off signalling with a low probability of the On signal.

\item \textit{Algorithmic PP-SWIPT modulations:} Inspired by the results obtained via learning,  an algorithmic approach is proposed for PP-SWIPT modulation design. The proposed algorithm performs very close to the performance corresponding to the modulations obtained via learning. In addition, unlike the modulation results obtained via learning (which requires a demanding training time for new system parameters), it has higher adaptivity towards the real-time change of system design parameters. The approach in this work (namely, the design of non learning-based algorithms inspired by the learning-based results) is novel and potentially can be extended to many other communications scenarios.

\item \textit{Learning multi-user SWIPT modulation  and coded modulation for PP-SWIPT:} The learning approach for PP-SWIPT modulation design is extended in two directions. First, different multi-user scenarios (Broadcast Channel, Multiple Access Channel and Interference Channel) are considered and different properties of the obtained results are extracted that can be potentially used to design algorithmic modulations similar to what is proposed for PP-SWIPT in this paper.  Second, coded modulation for PP-SWIPT is considered where the length of the channel input (corresponding to each transmit message) is larger than one. In particular, it is observed that by allowing the channel input length to be larger than one, it is feasible to reduce the \textit{Symbol Error Rate} (SER) while keeping the delivered power high.

\item \textit{Algorithmic coded modulation design for PP-SWIPT:} Inspired by the results obtained via learning for coded modulation, an algorithmic approach is proposed. The proposed algorithm is superior to the learning counterpart, in the sense that first, it significantly reduces the time and energy burden required to obtain the appropriate channel input for new system parameters (high adaptivity). Second, for relatively longer channel inputs, where learning is not feasible (due to curse of dimensionality), the algorithmic approach extends successfully and performs well (extendability to larger blocklengths). To the best of our knowledge, this is one of the only examples where Machine Learning (ML) is utilized to design algorithmic solutions.

\end{itemize}
The rest of the paper is organized as follows. In Section \ref{Sec:sys_Model}, the system model is introduced. In Section \ref{Sec:EH_Models}, after reviewing EH models proposed in the literature, a parametric model for an EH is obtained utilizing real data corresponding to input-power/output-power of an EH. In Section \ref{Sec:Modulation_Design_Learning}, the learning approach is used to study modulation design for PP-SWIPT. Inspired by the results obtained via the learning approach, an algorithmic PP-SWIPT modulation design is proposed in Section \ref{Sec:Modulation_Design_Algorithmic}. In Section \ref{Sec:Extension to Multi-user SWIPT}, extension of the learning approach to multi-user SWIPT is considered. Extending the results to coded modulation in Section \ref{Sec:Extension_to_coded_modulation}, we first study the learning approach in Section \ref{Sec:Coded_Design_Learning}. Next, in Section \ref{Sec:Coded_Design_Algorithmic}, an algorithm is proposed for designing coded modulation for PP-SWIPT. The paper is concluded in Section \ref{Sec:Conclusion}.

%%%%%%%%%%%%%%%%%%%%%%%%%%%%%%%%%%%%%%

\begin{figure}
\begin{centering}
\includegraphics[scale=0.38]{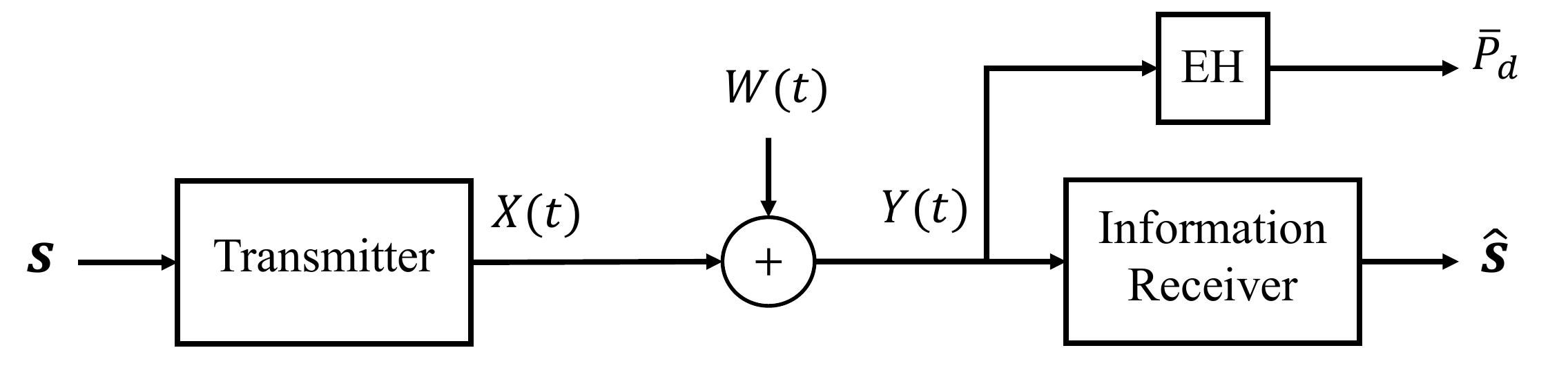}\vspace{-3mm}
\caption{PP-SWIPT system model}\label{Fig_sys_model}
\par\end{centering}
\vspace{-6mm}
\end{figure}

\vspace{-1mm}
\section{System Model}\label{Sec:sys_Model}

A PP-SWIPT system model is shown in Figure \ref{Fig_sys_model}. The transmitter communicates using $M$ possible messages $\pmb{s}\in \mathcal{M} = \{1,2,...,M\}$, where $\mathcal{M}$ denotes the message alphabet set. The mapping from the set of messages $\mathcal{M}$ to the transmitted information-power codeword $\pmb{x}^n$ is denoted by $g_{\theta_T}(\cdot): \mathcal{M} \rightarrow \mathbb{C}^n$, where $\theta_T$ refers to the set of transmitter design parameters. The transmitted RF signal is modelled as $X_{\text{RF}}(t)=\sqrt{2}\Re\{X(t)e^{j2\pi f_c t}\}$, where $X(t)$ is the baseband information-power bearing pulse modulated signal and $f_c$ is the carrier frequency. The baseband transmitted signal $X(t)$ is represented as $X(t) = \sum_{k=-\infty}^{\infty}\sum_{i=1}^{n}\pmb{x}_i(\pmb{s}) g(t-(kn+i)T)$, where the index $k$ represents the $k^{\text{th}}$ message in the transmitted sequence of messages, $g(t)$ is the pulse shaping waveform, and $\pmb{x}_i(\pmb{s})$ is the $i^{\text{th}}$ symbol of the complex information-power codeword $\pmb{x}^n=g_{\theta_T}(\pmb{s})$. We assume that the transmitted complex information-power symbols $x_i(s),~i=1,\ldots,n$ and $s=1,\ldots,M$ are under an average power constraint, i.e., we have
\begin{align}\label{Eq_11}
    \frac{1}{Mn}\sum_{s=1}^{M}\sum_{i=1}^{n}|x_i(s)|^2\leq P_a.
\end{align}

The receiver is assumed to be jointly capable of harvesting the energy of the received RF signal (denoted as $Y_{\text{RF}}(t)$) as well as estimating the transmitted message $\pmb{s}$\footnote{The assumption of joint harvesting-decoding is an ideal scenario and is made for simplicity. The approach can be extended to other types of receivers, e.g. power splitting, time switching or scenarios where the EH and ID receiver are not co-located.}. The EH is fed with the received RF signal, i.e., $Y_{\text{RF}}(t)=\sqrt{2}\Re\{Y(t)e^{j2\pi f_c t}\}$, where $Y(t)$ is the received signal in the baseband (from which the transmitted message is estimated) and is modelled as $Y(t) = X(t)+W(t)$ with $W(t)$ being the baseband complex-valued noise signal\footnote{The AWGN assumption is due to the fact that Gaussian noise is the worst noise for estimation.}. The expected delivered power corresponding to message $s\in \mathcal{M}$ is denoted as $P_{d}(s)$ (the expectation is taken with respect to the AWGN). The estimated message $\hat{\pmb{s}}$ is obtained by processing $\pmb{y}^n$ via the information receiver, where $\pmb{y}^n$ is the n-length complex samples of the signal $Y(t)$ taken with frequency $1/T$ Hz. The estimation is performed by mapping the received noisy codeword $\pmb{y}^n$ to $\hat{\pmb{s}}\in\mathcal{M}$ through a parametric function denoted by $h_{\theta_R}(\cdot): \mathbb{C}^n \rightarrow \mathcal{M}$, where $\theta_R$ refers to the set of receiver design parameters across the information receiver module.

We assume that the message set $\mathcal{M}$ is finite, discrete and possesses a uniform distribution over its support. For a given set of message set $\mathcal{M}$ at the transmitter, average power constraint $P_a$, and transmitted code length $n$, the goal is to design the channel inputs $\pmb{x}^n$, such that the receiver demand of information-power is (completely or at least partially) satisfied. Equivalently, we aim at finding a set of design parameters, that is $(\theta_T^{*},\theta_R^{*})$, such that the objective function $L(\theta_T^{*},\theta_R^{*})$ is minimized, where $L(\theta_T,\theta_R)$ is defined as
\begin{align}\label{Eq_6}
    L(\theta_T,\theta_R)=\mathbb{E}\left[-\log{p_{\hat{\pmb{s}}|\pmb{s}}}(\hat{\pmb{s}}=\pmb{s}|\pmb{s})+\frac{\lambda}{P_d(\pmb{s})}\right],
\end{align}
where the expectation is taken over the randomness of the system (AWGN and the message set). Throughout the paper, $\bar{P_d}$ denotes the average delivered power, i.e., $\bar{P_d}=\mathbb{E}[P_d(\pmb{s})]$.

\vspace{-1mm}
\begin{rem}
The motivation behind defining the objective function in the form of (\ref{Eq_6}) is that, first, probability of error, i.e., $P_e = \textrm{Pr}\{\hat{\pmb{s}}\neq \pmb{s}\} $, can be upperbounded by the term $\mathbb{E}[-\log{p_{\hat{\pmb{s}}|\pmb{s}}}(\hat{\pmb{s}}=\pmb{s}|\pmb{s})]$. Accordingly, minimizing the latter guarantees an upper bound on the error probability $P_e$. Second, to keep the objective function positive, we consider the delivered power in the denominator in (\ref{Eq_6}), i.e., $\frac{\lambda}{P_d(\pmb{s})}$. However, other alternatives can be chosen in order to take the delivered power into account.
\end{rem}
\vspace{-1mm}

%%%%%%%%%%%%%%%%%%%%%%%%%%%%%%%%%%%%%%

\begin{figure}
\begin{centering}
\includegraphics[scale=0.5]{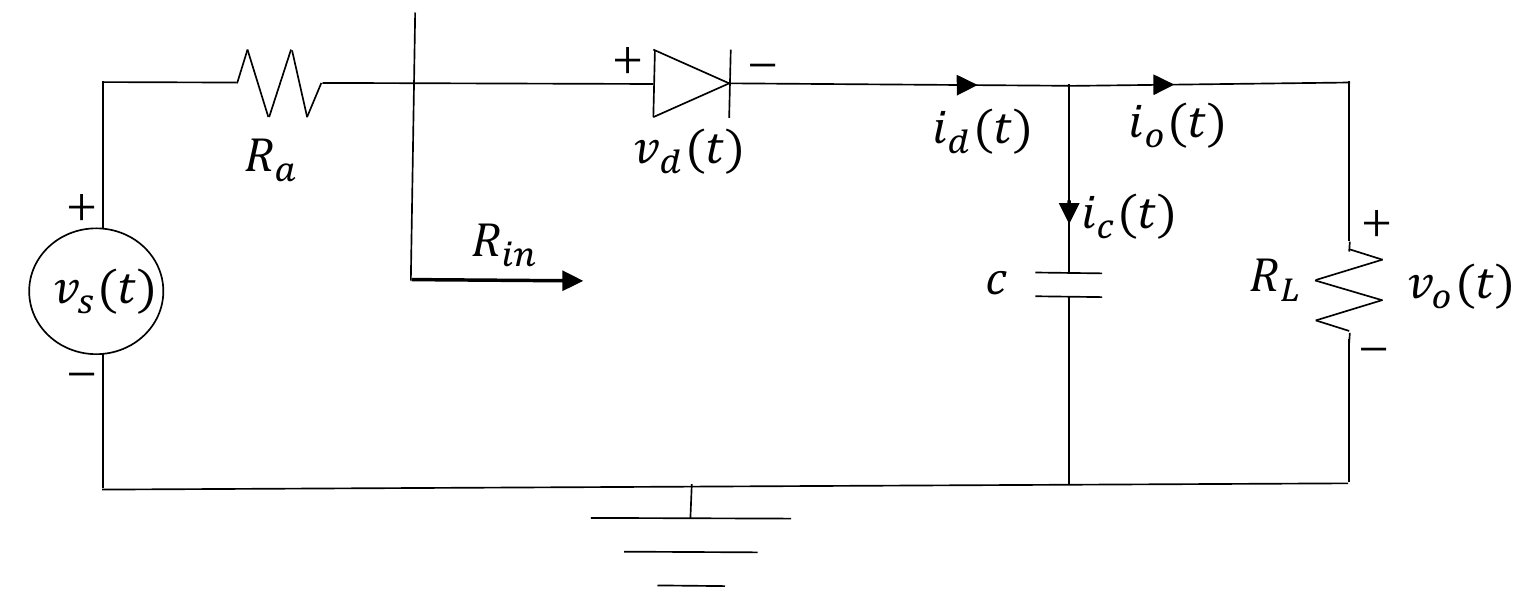}\vspace{-3mm}
\caption{Model of a practical EH}\label{Fig_rectenna}
\par\end{centering}
\vspace{-6mm}
\end{figure}

\vspace{-1mm}
\section{EH Models}\label{Sec:EH_Models}
A simple structure of an EH, namely a rectenna, is illustrated in Figure \ref{Fig_rectenna}. The received RF signal $Y_{\text{RF}}(t)$ is converted at the rectifier's output into a DC signal across a load resistance $R_L$. In the literature, depending on the application and available resources, different analytical models of an EH have been proposed accounting for practical nonlinear behaviour of an EH, such as nonlinear relationship between an EH's input- and output-power. These models can be categorized into two groups, namely, low and high EH input power regimes. The main focus of the rectifier in the low-power regime (about $-30$dBm to $-5$dBm input power) is to operate in the square law and transition zones \cite{Boaventura_Collado_Carvalho,Clerckx_Zhang_Schober_Wing_Kim_Vincent}, which also captures the \textit{threshold effect}, i.e., the power level below which the rectifier turns off. However, for the models suitable for high-power regime (above $-5$dBm input power), the main focus is to capture the \textit{saturation effect}, i.e., the power level above which the rectifier's output DC power level remains constant.\footnote{Operating a diode in the breakdown region is not the purpose of a rectifier and should be avoided as much as possible \cite{Boaventura_Collado_Carvalho,Clerckx_2016,Clerckx_Zhang_Schober_Wing_Kim_Vincent}.}

In the following, we first review existing EH models in Sections \ref{Sec:Low_Power} and \ref{Sec:high_Power} for low and high EH input power regimes, respectively. Next in Section \ref{Sec:Learned_Model}, we propose a parametric model obtained by applying a non-linear regression algorithm over collected real data. Note that an EH harvests the energy of a signal in an instantaneous manner, that is, without requiring to store/process the whole transmitted block of signal. Accordingly, for brevity in this section, we consider the channel input $\pmb{x}^n$ to be of length one, i.e., $n=1$. It is straightforward to verify that the harvested energy corresponding to a channel input of length $n$ is the summation of the harvested energy via each symbol of the transmitted codeword $\pmb{x}^n$.

%%%%%%%%%%%%%%%%%%%%%%%%%%%%%%%%%%%%%%
\vspace{-1mm}

\subsection{Low EH Input Power Regime}\label{Sec:Low_Power}
\begin{itemize}
\item \textit{Model $A$}:
In \cite{Clerckx_Bayguzina_2016}, an EH model based on the Taylor expansion of the diode characteristic function is introduced. It is shown that the delivered power is a function of even moments of the received RF signal $Y_{\text{RF}}(t)$ \cite[eq. 7]{Clerckx_Bayguzina_2016}, which can be approximated by truncating it up to the fourth moment \cite[eq. 25]{Clerckx_Bayguzina_2016}. In \cite[Prop. 3]{Varasteh_Rassouli_Clerckx_arxiv}, it is shown that the delivered power for \textit{independent and identically distributed} (iid) inputs is a function of the quantities $\mathbb{E}[|\pmb{x}|^4]$ and $\mathbb{E}[|\pmb{x}|^2]$, as well as first to fourth moments of the real and imaginary components of the channel input $\pmb{x}$ (for more detail see \cite{Varasteh_Rassouli_Clerckx_arxiv}).

\item \textit{Model $B$}:
In \cite{Vedady_Zeng}, an EH model based on the Bessel function representation of the diode characteristic function is introduced. Adopting the rectenna model in \cite{Clerckx_Bayguzina_2016}, the received RF signal $Y_{\text{RF}}(t)$ is converted at the rectifier's output into a DC signal across a load resistance $R_L$. Assuming that the capacitance $c$ is sufficiently large, the output voltage is approximately constant, i.e., $v_{o}(t) \thickapprox v_{o}$. Additionally, assuming perfect impedance matching and applying Kirchoff's current law to the circuit in Figure \ref{Fig_rectenna}, it is shown in \cite{Varasteh_Piovano_Clerckx} that the delivered power $\bar{P_d}$ is obtained as
$\bar{P_d} = f^{-1}\left(\mathbb{E}\left[I_0(\sqrt{2}B|\pmb{x}|)\right]\right)$, where $B$ is a constant, and $f(x)=\left(1+\frac{\sqrt{x}}{i_s\sqrt{R_L}}\right)\exp\left(B\sqrt{\frac{R_L x}{R_a}}\right)$.
\end{itemize}
%$B = \frac{\sqrt{R_a}}{\eta V_T}$

\vspace{-1mm}
\subsection{High EH Input Power Regime}\label{Sec:high_Power}
\begin{itemize}
\item \textit{Model $C$}:
There are diminishing returns and limitations on the maximum possible harvested power whenever the same rectifier is used for the high-power regime \cite{Boaventura_Collado_Carvalho,Valenta_Durgin}.\footnote{Adapting the rectifier as the input power level increases would avoid the saturation problem \cite{Boaventura_Collado_Carvalho}.} In \cite{Boshkovska}, a nonlinear model based on the sigmoidal function is proposed that captures this practical limitation. The delivered power $\bar{P_d}$ is modelled as $\bar{P_d}=\mathbb{E}\left[\frac{\Psi(\pmb{y})-L_s\Omega}{1-\Omega}\right]$, where $\Omega=1/(1+ \exp(a b))$ and $\Psi(\pmb{y})=L_s/(1+\exp(-a (|\pmb{y}|^2-b)))$. The parameter $\Omega$ is to ensure the zero-input/zero-output response for the EH, and the parameter $L_s$ denotes the maximum harvested power when an EH is saturated\footnote{We note that model $C$ is based on fitting for a given input signal shape and expresses the dependency only on the input signal power, whereas models $A$ and $B$ express the dependency on the entire input signal, not only its power.}.
\end{itemize}

%%%%%%%%%%%%%%%%%%%%%%%%%%%%%%%%%%%%%%

\vspace{-1mm}

\subsection{Parametric Model Based on Learning from Data}\label{Sec:Learned_Model}
The existing EH models in the literature (as reviewed earlier) are obtained based on approximations and simplifying assumptions. Accordingly, any approach towards signal design (based on the reviewed EH models) does not feature a real and unified layout for the studied problem. In particular, it is not known how to design the information-power bearing signal in transition regime (from -15 to -5 dBm). Additionally, as shown later, the low-power models A and B, yield slightly different solutions.

Unfortunately, analytic modelling of a practical EH has shown extremely demanding due to the many obstacles we face in modelling the different elements in an EH, such as the diode nonlinear characteristic function, impedance mismatch of the antenna, non-ideal low pass filter and so on. Although using the basic laws in circuit theory, some accurate models (without approximations and simplifying assumptions) can be obtained, those models are too cumbersome to be tackled analytically.

\begin{figure}[!tbp]
  \centering
  \begin{minipage}[b]{0.4\textwidth}
    \includegraphics[width=\textwidth]{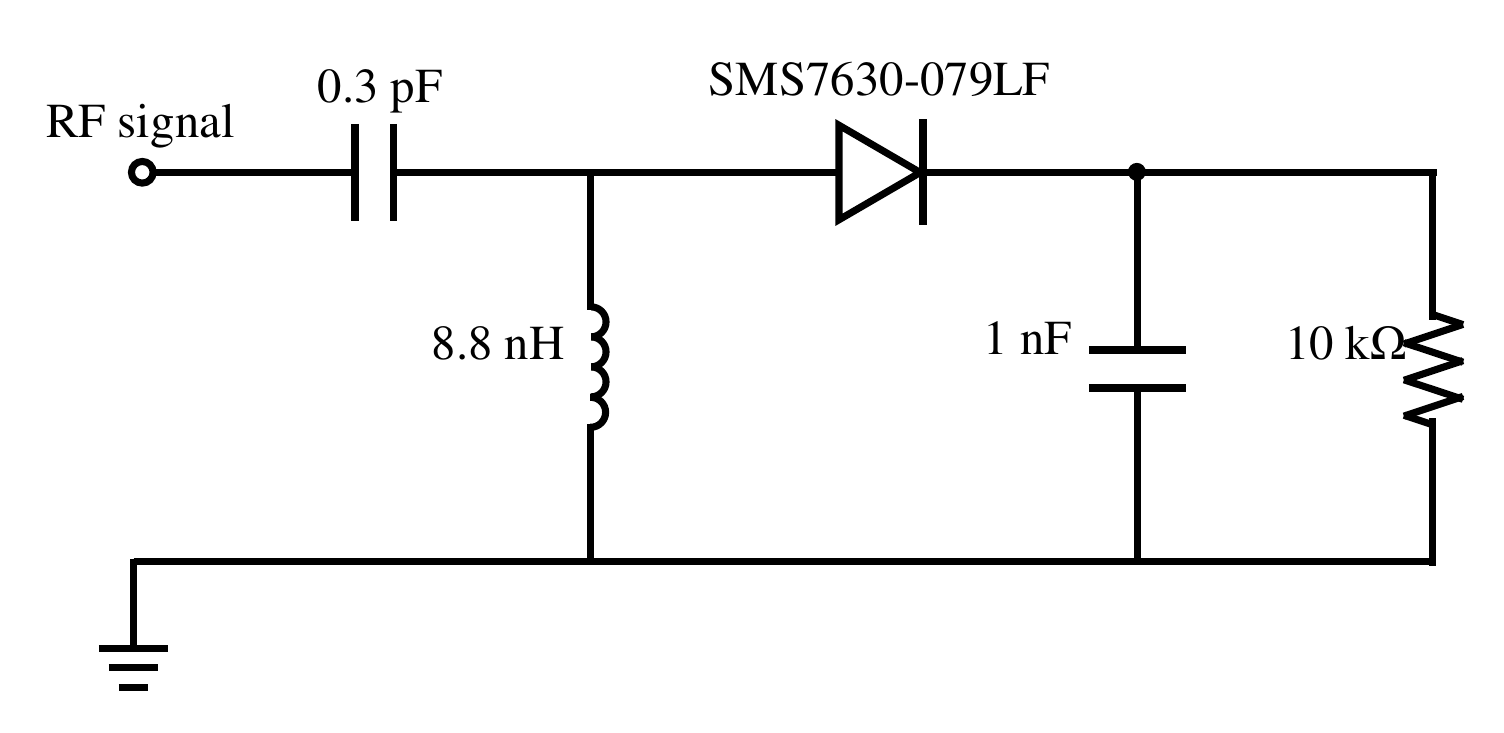}\vspace{-3mm}
    \caption{The EH circuit in \cite{Kim_Clerckx} based on which the measurements are collected. Collected data is used to learn the parametric model of the EH.}\label{Fig_circuit_Junghoon}
  \end{minipage}
  \hfill
  \begin{minipage}[b]{0.44\textwidth}
    \includegraphics[width=\textwidth]{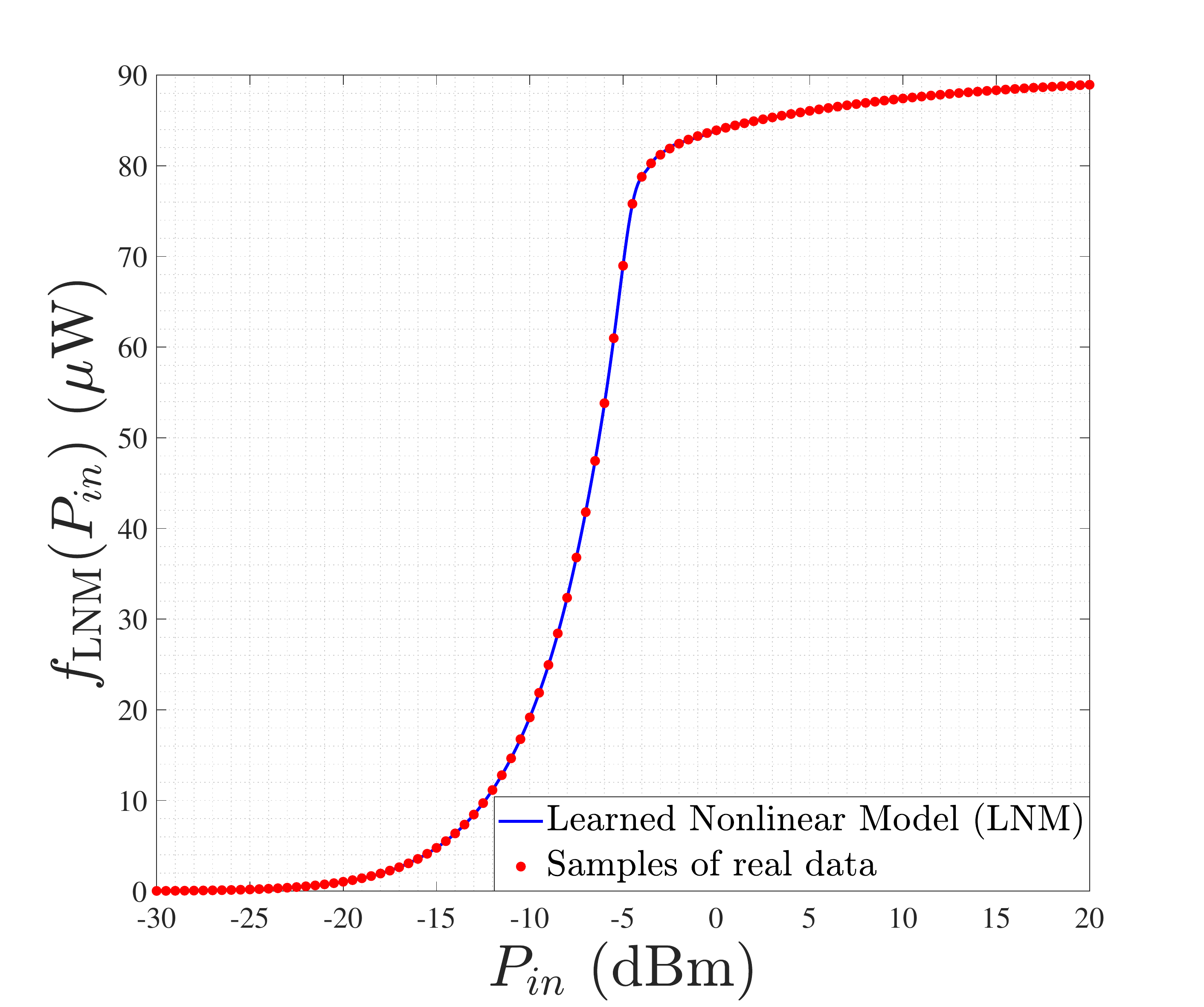}\vspace{-3mm}
    \caption{Model of an EH based on applying nonlinear regression over collected real data.}\label{Fig_LNM}
  \end{minipage}
  \vspace{-6mm}
\end{figure}

One promising alternative to EH modelling is to learn its input-power/output-power relationship from measurements. The larger the amount and diversity (in terms of a large range of input power, load, etc) of the collected data, the higher the accuracy of the learned model. In the following, we obtain an EH model by applying nonlinear regression over collected real data. In particular, we study the data collected from the EH circuit with specifications illustrated in Figure \ref{Fig_circuit_Junghoon}. The function we consider for modelling the EH is given as
\begin{align}\label{Eq_5}
    f_{\text{LNM}}(P_{\text{in}}) = \sigma(W_3\sigma(W_2\sigma(W_1 P_{\text{in}} +w_1)+w_2)+w_3),
\end{align}
where $P_{\text{in}}$ denotes EH instantaneous input power, i.e., $P_{\text{in}}=|\pmb{y}|^2$. $\mathcal{W}=\{W_1^{3\times 1},$ $W_2^{2\times 3},$ $W_3^{1\times 1},$ $w_1^{3\times 1},$ $w_2^{2\times 1},$ $w_3^{1\times 1}\}$ is the set of parameters to be optimized and $\sigma(\cdot)=\tanh(\cdot)$. In order to train the parameters we use Gradient Descent optimization applied over the following objective function
\begin{align}
    L_{\text{EH}}(\mathcal{W}) = \frac{1}{m}\sum_{i=1}^{m} (f_{\text{LNM}}(P_{\text{in}})-P_{\text{out}})^2,
\end{align}
where $P_{\text{out}}$ is the collected output power corresponding to an input power $P_{\text{in}}$ and $m$ is the number of collected data used for training. In Figure \ref{Fig_LNM}, the learned model (solid blue line) and the collected data (red dots) are illustrated\footnote{The measurements are obtained using Continuous Wave (CW) signals and assuming that the circuit is operating in steady state (since the effect of the transient state is negligible). Modulation learning using CW measurements is justified because amplitude modulation effectively corresponds to a CW with different power levels.}. Throughout this paper, we adopt the model introduced here.

%\begin{figure}
%\begin{centering}
%\includegraphics[scale=0.43]{simulation_rectifier.pdf}
%\caption{The EH circuit based on which the real data corresponding to input-output power is collected. The collected data is used via nonlinear regression algorithm to learn an analytical model \cite{Kim_Clerckx}.}\label{Fig_circuit_Junghoon}
%\par\end{centering}
%\vspace{-5mm}
%\end{figure}
%
%\begin{figure}
%\begin{centering}
%\includegraphics[scale=0.21]{LNM.pdf}
%\caption{Model of a practical EH based on applying nonlinear regression over collected real data.  {\color{Jakob}Jakob: About the transient effect, we assume the rectifier operates in the steady state mode. }}\label{Fig_LNM}
%\par\end{centering}
%\vspace{-5mm}
%\end{figure}

\subsubsection{Observations on the Proposed Parametric Model in (\ref{Eq_5})}\label{Sec:observations}
Based on the proposed model in (\ref{Eq_5}), here we analyze the behaviour of the EH in more details, which leads to some interesting observations. These observations will come useful in explaining the results obtained via learning.

In \cite{Varasteh_Rassouli_Clerckx_arxiv}, the capacity of a point-to-point channel with a nonlinear EH at the receiver is studied. It is observed that as the receiver power demand increases, the capacity achieving inputs approach to On-Off keying signalling where the On signal has a high-amplitude/low-probability. Inspired by this result, here (by focusing only on the power delivery perspective) we assume that the channel input $\pmb{x}\triangleq\pmb{r}e^{j\pmb{\theta}}$ follows a \textit{probability density function} (pdf) given as
\begin{align}\label{Eq_9}
 p_{\pmb{r},\pmb{\theta}}(r,\theta)=\frac{1}{2\pi}((1-p_{on})\delta(r)+p_{on}\delta(r-\sqrt{P_a/p_{on}})),
\end{align}
where $p_{on}$ is the probability of the ON signal and $\delta(\cdot)$ is the Dirichlet function. Note that with an input as in (\ref{Eq_9}), the channel average power constraint is $\mathbb{E}[|\pmb{x}|^2]= P_a$.

\begin{figure}[!tbp]
  \centering
  \begin{minipage}[b]{0.44\textwidth}
    \includegraphics[width=\textwidth]{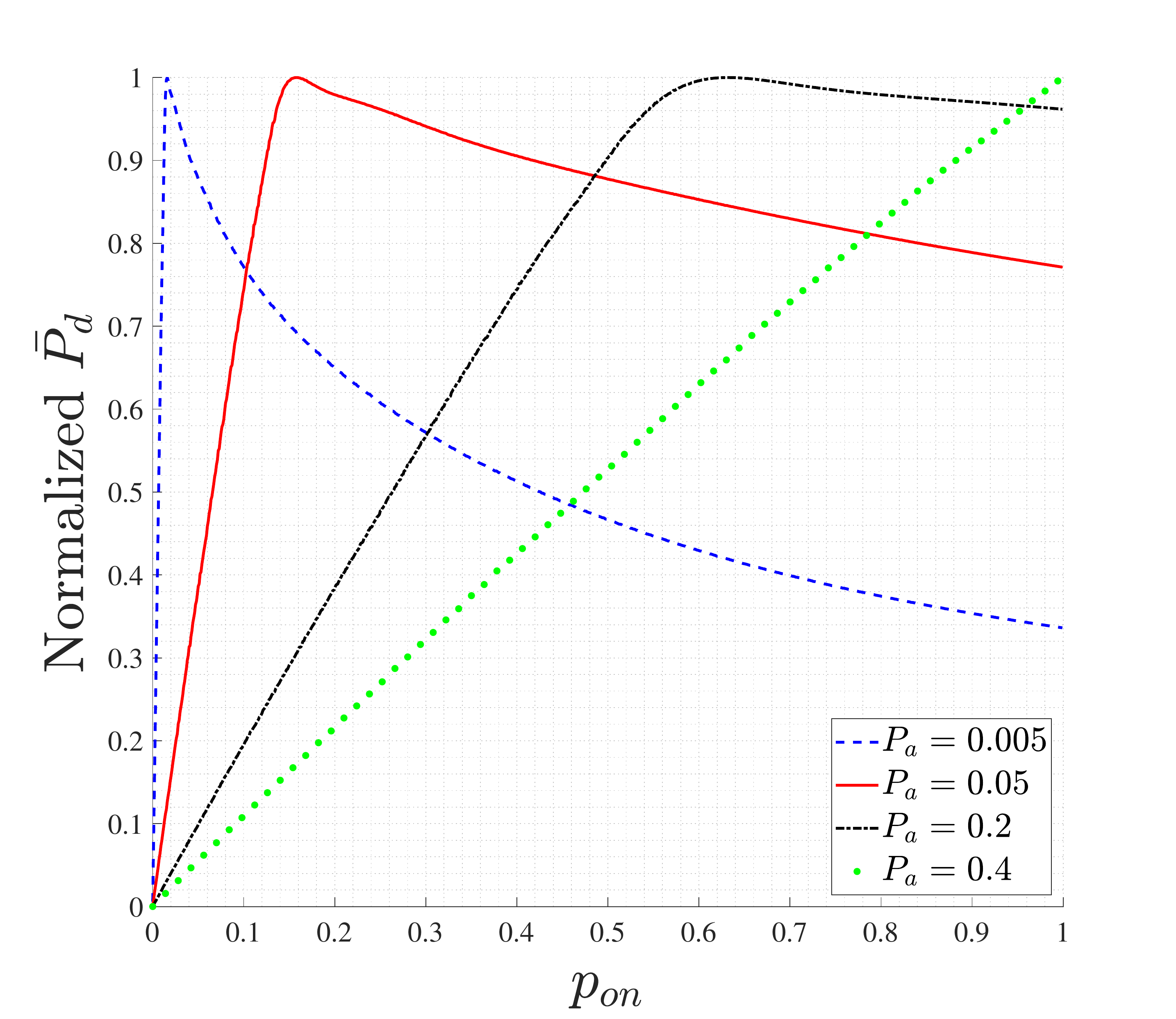}\vspace{-3mm}
    \caption{Normalized delivered power $\bar{P}_d$ of the model in (\ref{Eq_5}) under different average power constraints with channel input given in (\ref{Eq_9})}\label{Fig_Normalized_DelPow}
  \end{minipage}
  \hfill
  \begin{minipage}[b]{0.44\textwidth}
    \includegraphics[width=\textwidth]{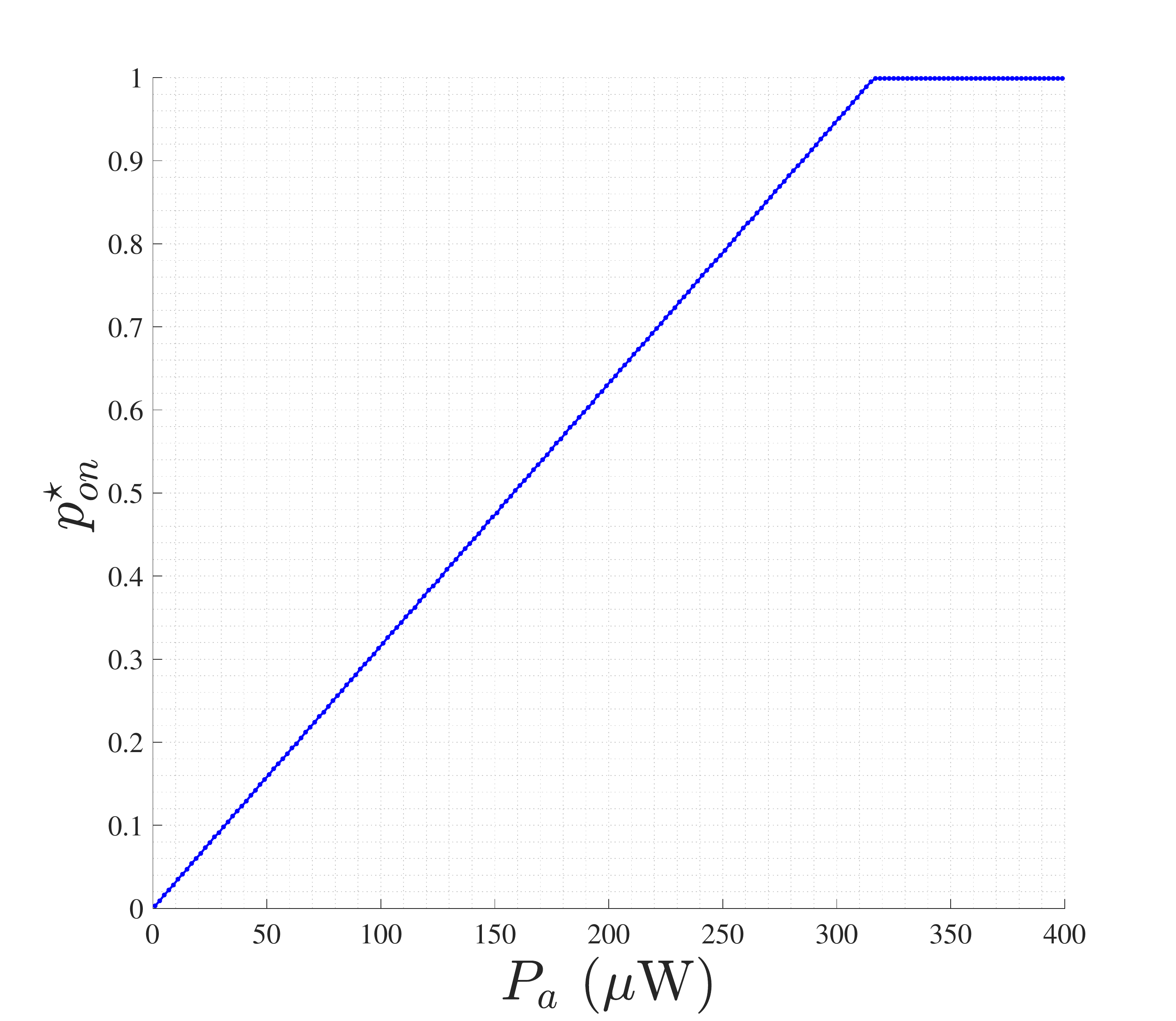}\vspace{-3mm}
    \caption{Optimal value for the probability of the ON signal $p_{on}^{*}$ in (\ref{Eq_9}) versus the channel average power constraint $P_a$}\label{Fig_Optima_Probability}
  \end{minipage}
  \vspace{-6mm}
\end{figure}

%\begin{figure}
%\begin{centering}
%\includegraphics[scale=0.22]{Pd_vs_prob.pdf}
%\caption{Normalized average delivered power $\bar{P}_d$ corresponding to the model in (\ref{Eq_5}) under different average power constraints and with the channel input following the distribution given in (\ref{Eq_9})}\label{Fig_Normalized_DelPow}
%\par\end{centering}
%\vspace{-5mm}
%\end{figure}
%
%\begin{figure}
%\begin{centering}
%\includegraphics[scale=0.22]{Optimal_Prob_vs_Power.pdf}
%\caption{Optimal value for the probability of the ON signal $p_{on}$ in (\ref{Eq_9}) versus the channel average power constraint $P_a$.}\label{Fig_Optima_Probability}
%\par\end{centering}
%\vspace{-5mm}
%\end{figure}

In Figure \ref{Fig_Normalized_DelPow}, the normalized average delivered power\footnote{The normalized average delivered power is given as $\bar{P}_{d,\text{normalized}}=\frac{\mathbb{E}[f_{\text{LNM}}(|\pmb{y}|^2)]}{\bar{P}_{d,\text{max}}}$, where $\bar{P}_{d,\text{max}}$ is the maximum delivered power under the average power constraint of $P_a$.} for inputs of the form (\ref{Eq_9}) and for different values of the probability of the On signal $p_{on}$ is illustrated under four different average power constraints.\footnote{In the simulations, for brevity, we have assumed a noiseless channel, i.e., we have $\bar{P}_{d,\text{normalized}}=\frac{\mathbb{E}[f_{\text{LNM}}(\pmb{r}^2)]}{\bar{P}_{d,\text{max}}}=\frac{p_{on} f_{\text{LNM}}(P_a/p_{on})}{p_{on}^{\star} f_{\text{LNM}}(P_a/p_{on}^{\star})}$.} It is observed that for power transmission purposes, there is only one value for $p_{on}$, which results in the maximum delivery power. In Figure \ref{Fig_Optima_Probability}, the optimal value of $p_{on}$, denoted by $p_{on}^{\star}$, (the probability which results in the maximum delivered power) for different average power constraint values is plotted. It is observed that as the average power constraint increases, the probability $p_{on}^{\star}$ increases as well. This increase in the probability $p_{on}^{\star}$, continues until we have $p_{on}^{\star}=1$ for average power constraints (roughly) above $317~\mu$W. That is for average power constraint higher than $317~\mu$W, constant amplitude signalling with amplitude $\pmb{r} = \sqrt{P_a}$  is suitable for power delivery purposes. The probability $p_{on}^{\star}$ as a function of the EH input power can be approximated as
\begin{align}\label{Eq_13}
 p_{on}^{\star}=\min\{ P_a/317,1\},
\end{align}
where $P_a$ is with unit $\mu$W.

\vspace{-2mm}
\section{Modulation Design for poinit-to-point SWIPT}\label{Sec:SWIPT_PP_Modulation_Design}
In this section, we consider two approaches to design modulations for PP-SWIPT. First, in Section \ref{Sec:Modulation_Design_Algorithmic}, we follow a learning approach by considering the system in Figure \ref{Fig_sys_model} as an NN-based denoising autoencoder. Inspired by the learning results in Section \ref{Sec:Modulation_Design_Learning}, in Section \ref{Sec:Modulation_Design_Algorithmic}, we propose an algorithmic modulation design for PP-SWIPT. In Section \ref{Sec:Modulation_Design_Performance}, the performance of the designed modulations are provided and compared.

\begin{figure}
\begin{centering}
\includegraphics[scale=0.42]{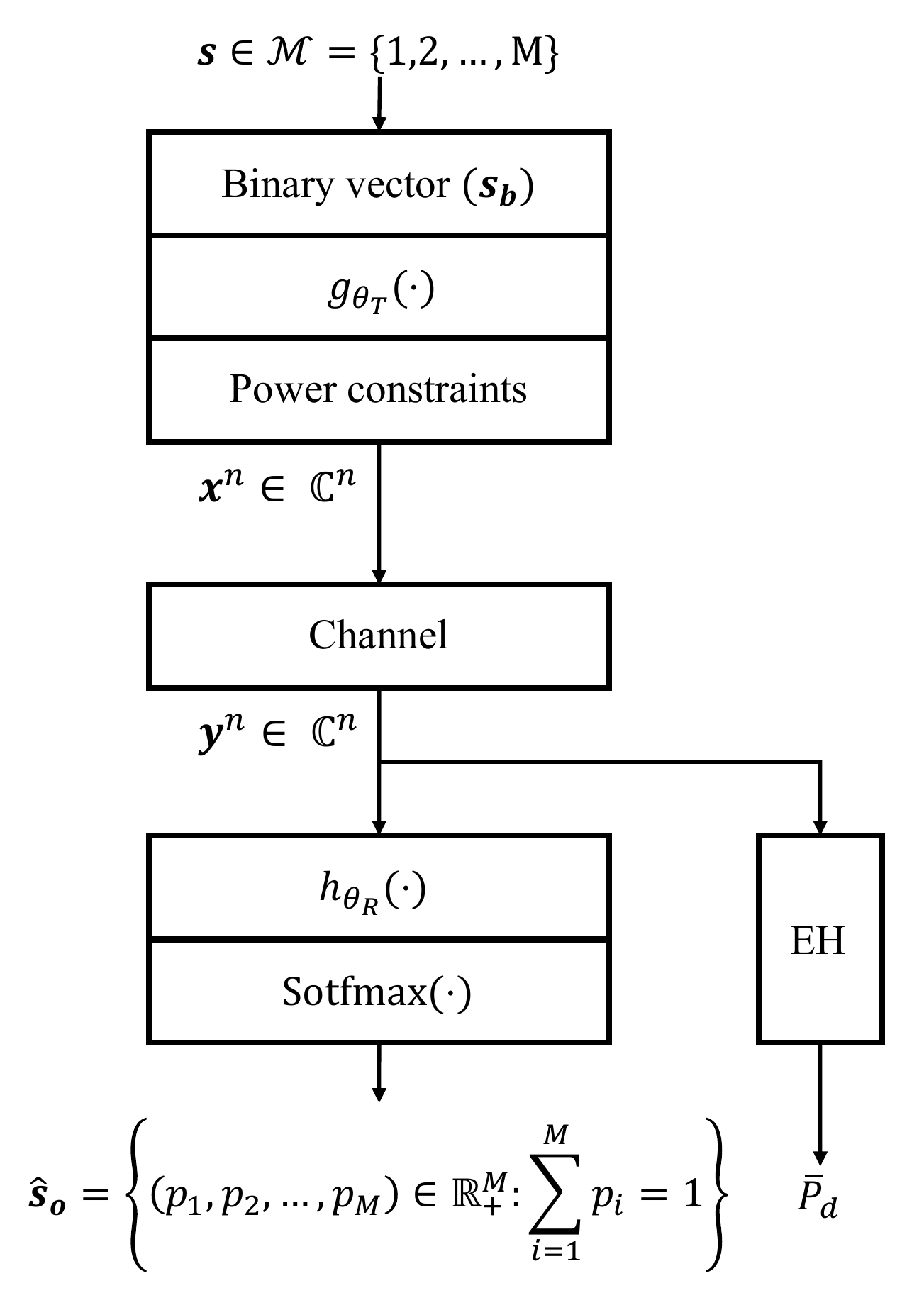}\vspace{-3mm}
\caption{NN-based implementation of the system in Figure \ref{Fig_sys_model}}\label{Fig_nn_pp_structure}
\par\end{centering}
\vspace{-8mm}
\end{figure}

\vspace{-1mm}
\subsection{Learning Approach}\label{Sec:Modulation_Design_Learning}
We consider the PP-SWIPT system in Figure \ref{Fig_sys_model} as an NN-based denoising autoencoder, where both the transmitter and receiver are implemented as NNs in order to perform the modulation and demodulation processes, respectively. A general NN-based implementation of the system in Figure \ref{Fig_sys_model} is illustrated in Figure \ref{Fig_nn_pp_structure}. As shown in Figure \ref{Fig_nn_pp_structure}, at the transmitter, the message $\pmb{s} \in \mathcal{M}$ is converted into a binary vector of length $\lceil\log_{2}(M)\rceil$ denoted by $\pmb{s_b}$ ($\lceil x \rceil$ returns the smallest integer larger than $x$). The vector $\pmb{s_b}$ is then processed by the NN and is converted into a codeword $\pmb{x}^{n}=f_{\theta_{T}}(\pmb{s})$. Accordingly, the set of transmitter design parameters $\theta_T$ are related to the weights and biases across the encoder module. To satisfy the power constraint, a power normalization is performed as the last layer of the transmitter module.

The codeword $\pmb{x}^n$ is corrupted by the channel noise. At the receiver, the estimation is performed by mapping the received noisy codeword $\pmb{y}^n$ to an $M$-dimensional output probability vector denoted by $\pmb{\hat{s}_o}$ (and estimating the message by returning the index corresponding to the maximum probability). Accordingly, $\theta_R$ refers to the set of receiver parameters in terms the weights and biases across the decoder module.

The delivered power at the receiver is modelled as in (\ref{Eq_5}). Note that, since for power delivery purposes, the received RF signal is directly fed into the EH, the signal is not processed through the NN. We aim at minimizing the objective function in (\ref{Eq_6}) by following a learning approach and training the structure in Figure \ref{Fig_nn_pp_structure}. Accordingly, the objective function $L(\theta_T,\theta_R)$ in (\ref{Eq_6}) is approximated by $L_a(\theta_T,\theta_R)$ given as
\begin{align} \label{Eq_4}
L_a(\theta_T,\theta_R) = \frac{1}{|\mathcal{B}|}\sum_{l\in \mathcal{B}} \mathcal{L}(s_{o}^{(l)}, \hat{s}_o^{(l)})  +\frac{\lambda}{P_d(s^{(l)})},
\end{align}
where $\mathcal{B}$ is a randomly drawn minibatch of training data, which is assumed to be generated iid with a uniform distribution over the message set $\mathcal{M}$. $\mathcal{L}(s_{o}^{(l)}, \hat{s}_o^{(l)}) = -\sum_{i=1}^M s_{o,i}^{(l)} \log {\hat{s}_{o,i}^{(l)}}$ is the cross entropy function between the one-hot representation of the $l^{th}$ training sample (denoted by $s_{o}^{(l)}$) and its corresponding output probability vector $\hat{s}_o^{(l)}$ at the receiver. $s_{o,i}^{(l)}$ and $\hat{s}_{o,i}^{(l)}$ indicate the $i^{\text{th}}$ entry of the vectors $s_o^{(l)}$ and $\hat{s}_o^{(l)}$, respectively.

\vspace{-1mm}
\begin{rem}
In NN-based implementation of the system in Figure \ref{Fig_sys_model}, the learning algorithm is performed over the baseband samples of the transmitted and received signal, that is $\pmb{x}^n$ and $\pmb{y}^n$, respectively. The validity of the approach is justified by noting that in (\ref{Eq_5}), the EH input power $P_{\text{in}}$ is the power of the RF process $Y_{\text{RF}}(t)$, which is equal to the power of its baseband version $Y(t)$ and that can be expressed in terms of samples' power of the signal $Y(t)$.
\end{rem}
\vspace{-1mm}

\begin{rem}
Different values of the parameter $\lambda\geq 0$ in (\ref{Eq_4}) can be associated with different information-power demands. Equivalently, from a ML perspective, the addition of the term $\frac{\lambda}{P_d(s^{(l)})}$ in (\ref{Eq_4}) induces a prior knowledge on the space of system design parameters, and it can be interpreted as a regularizer.
\end{rem}

\vspace{-1mm}
\begin{rem}
From learning perspective, differentiability of the EH model is crucial. Alternatively, a real system can be optimized directly without modelling, using gradient estimation techniques (see, e.g., \cite{Hoydis_Aoudia}).
\end{rem}
\vspace{-1mm}

\begin{figure}
\begin{centering}
\includegraphics[scale=0.45]{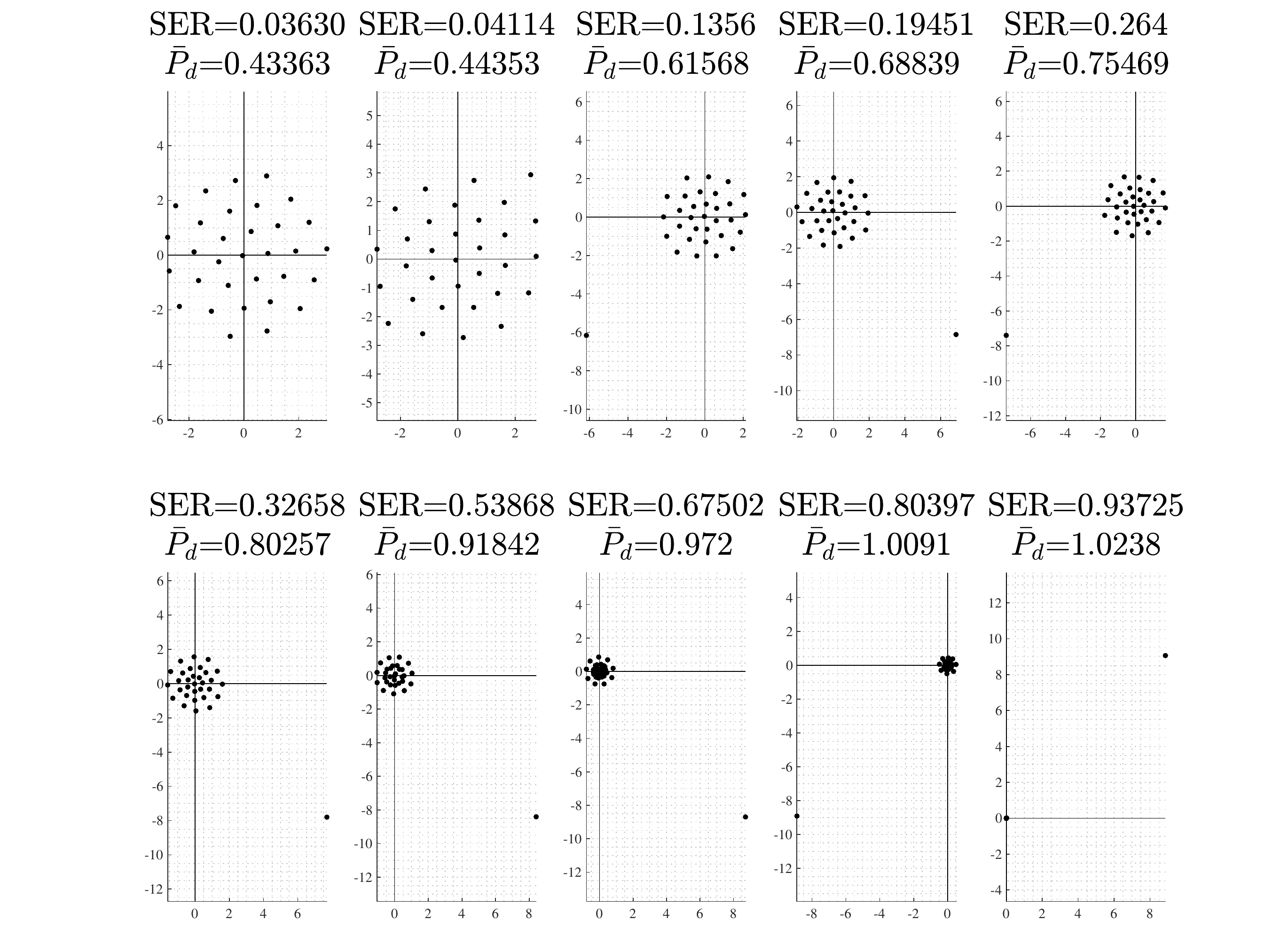}\vspace{-3mm}
\caption{Illustration of 32-symbols modulation for different values of $\lambda$ with SNR$=50$ ($16.98$ dB) and average power constraint $P_a=5~\mu$W. By increasing $\lambda$, the delivered power $\bar{P_d}$ at the receiver increases.}\label{Fig_Constellation_32_0p005}
\par\end{centering}
\vspace{-6mm}
\end{figure}

\begin{figure}
\begin{centering}
\includegraphics[scale=0.45]{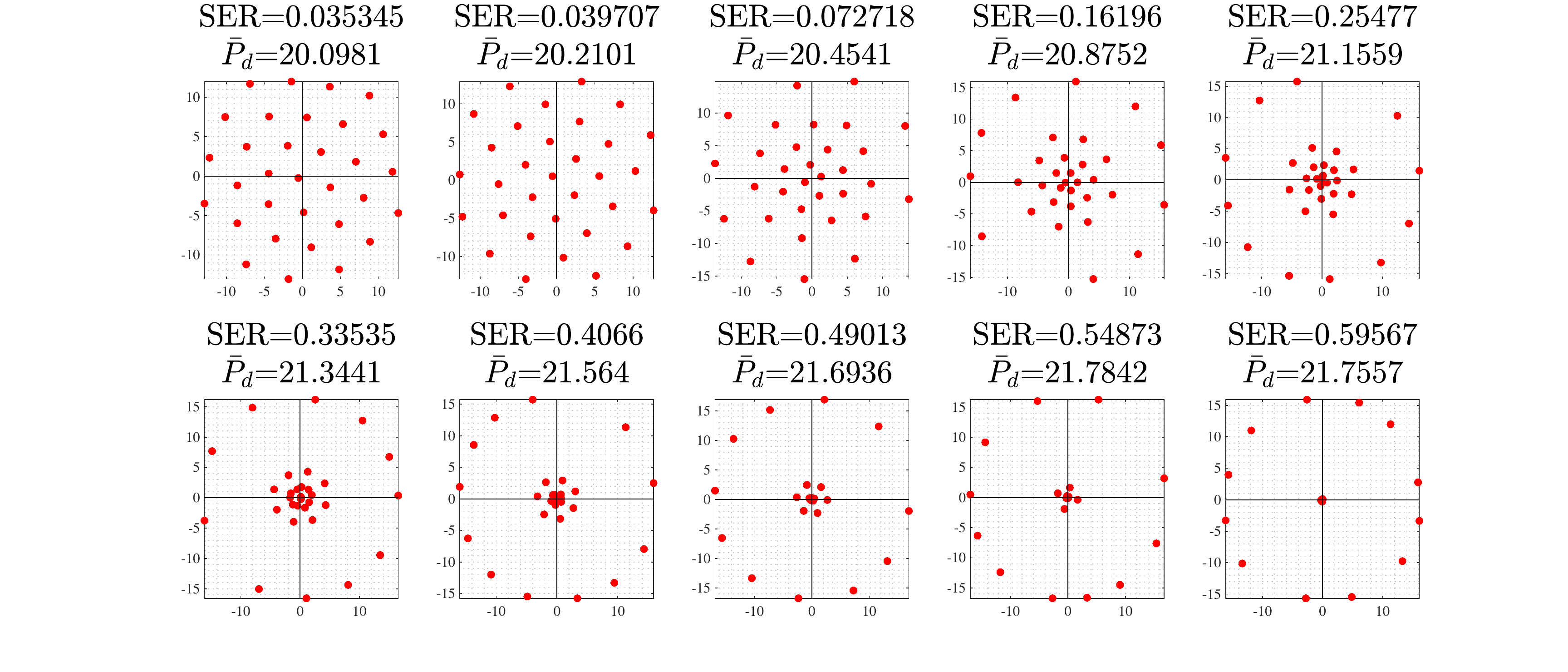}\vspace{-3mm}
\caption{Illustration of 32-symbols modulation for different values of $\lambda$ with SNR$=50$ ($16.98$ dB) and average power constraint $P_a=120~\mu$W. By increasing $\lambda$, the delivered power $\bar{P_d}$ at the receiver increases.}\label{Fig_Constellation_32_0p1}
\par\end{centering}
\vspace{-6mm}
\end{figure}

In Figures \ref{Fig_Constellation_32_0p005} and \ref{Fig_Constellation_32_0p1} the learned transmitted modulations are shown for message set size $M=32$ and for different values of $\lambda$ under average power constraints of $P_a = 5~\mu$W and $P_a=120~\mu$W, respectively, and with $\text{SNR}=50$ ($16.98$ dB). The observations are as follow:
\begin{itemize}
\item By increasing $\lambda$, the demand for power at the receiver increases. Accordingly, the modulation loses its symmetry around the origin in a way that some of the transmitted symbols (one symbol in Figure \ref{Fig_Constellation_32_0p005} and $12$ symbols in Figure \ref{Fig_Constellation_32_0p1}) are getting away from the origin. This can be explained as follows. It is verified from (\ref{Eq_13}) that for $P_a=5~\mu$W and $P_a=120~\mu$W, the optimal probabilities of the On signal in (\ref{Eq_9}) are $p_{on}^{\star}=5/317$ and $p_{on}^{\star}=120/317$, respectively. As the receiver power demand increases, the transmit signal modulation is reformed, such that it resembles an On-Off keying signalling with its corresponding $p_{on}^{\star}$. In the extreme scenario, where the receiver demand for power is at its maximum, the symbols possess only two amplitudes (one away from the origin and the other equal to zero). This is because with $M=32$ symbols, the closest On-Off keying signallings (to the ones with $p_{on}^{\star}=5/317$ and $p_{on}^{\star}=120/317$) are achieved by shooting one symbol (yielding the probability $1/32$ for the On signal) and $12$ symbols (yielding the probability $12/32$ for the On signal), respectively, away from zero amplitude.

\item An interesting observation about the PP-SWIPT modulations in Figure \ref{Fig_Constellation_32_0p005} (in particular focusing on the last modulations) is that, the channel input empirical distribution approaches to a distribution with two mass points for the amplitude, one with ``low-probability/high-amplitudes'' and the other with ``high-probability/zero-amplitudes''. This result is inline with \cite{Varasteh_Rassouli_Clerckx_arxiv}, where it is shown that the optimal channel input distributions for power delivery purposes (accounting for nonlinearity) follow the same behaviour, i.e., ``low-probability/high-amplitudes'' and ``high-probability/zero-amplitudes''.
\end{itemize}

\vspace{-1mm}

\begin{rem}
Similar to the scenario where the receiver can potentially harvest energy even if it demands merely information, it is possible to receive information when the receiver demands merely power. That is why the symbols on the outer circle in Figure \ref{Fig_Constellation_32_0p1} (the plots at the bottom) are placed equidistantly. Although the resulting SER for high-power demands is high (making estimation almost impossible), as it is discussed later (see Remark \ref{rem:1} in Section \ref{Sec:Modulation_Design_Performance}), the SER can be decreased by allowing block transmissions.
\end{rem}

\vspace{-1mm}
\begin{rem}
So long as the EH is not saturated, the harvested power increases by the number of messages $M$ (under the same average power constraint $P_a$). This is due to the fact that increasing the number of messages $M$, helps the transmitter approximate the optimal On-Off keying signalling more accurately.
\end{rem}

\vspace{-1mm}
\begin{rem}
The problem of learning signal modulations for PP-SWIPT is studied in \cite{Varasteh_Piovano_Clerckx} and \cite{Varasteh_Hoydis_Clerckx} under the assumption of model B and models A/model C, respectively. Although model A and model B are for EH low input power range, the learned modulations in \cite{Varasteh_Piovano_Clerckx} and \cite{Varasteh_Hoydis_Clerckx} result in slightly different ones. This highlights the need to learn a model which more accurately models the structure of an EH. The results obtained in this paper based on the proposed model in (\ref{Eq_5}) are inline with the results reported in \cite{Varasteh_Piovano_Clerckx} and \cite{Varasteh_Hoydis_Clerckx}. Moreover, the proposed model in (\ref{Eq_5}) provides a unified model for an EH, which in turn, allows the optimization to produce well performing solutions (in terms of transceiver design parameters $\theta_{T},\theta_{R}$) in the transition region (from $-15$ to $-5$ dBm) as well.
\end{rem}
\vspace{-1mm}

In Section \ref{Sec:Learned_Model}, it is observed that the maximum delivered power $\bar{P_d}$ is dependent on the probability of the On signal of the form (\ref{Eq_9}). Additionally, in Section \ref{Sec:Modulation_Design_Learning}, similar observations are made from the learned results, in the sense that the number of symbols shot away from zero increases with the average power constraint $P_a$. This is to maximize the delivered power $\bar{P_d}$ by approximating the probability $p_{on}^{\star}$ in (\ref{Eq_13}). Inspired by the aforementioned results, in the next section, we propose an analytical modulation design procedure that performs very close to the performance obtained from learning. The advantage of the analytical approach is that it does not require training. As a by-product, it can dynamically adapt to changes in the system design parameters. To the best of our knowledge, this is one of the only examples where ML helps to inspire an algorithmic solution.

\vspace{-1mm}
\subsection{Algorithmic Approach}\label{Sec:Modulation_Design_Algorithmic}
In the following, first, we explain the information modulation, and then, we explain how the information modulation can be transformed into PP-SWIPT modulations.

\subsubsection{Information Modulation}\label{Sec:Information_Only_Modulation}
Consider a constellation of $M$ complex points arranged over $C+1$ concentric circles with progressive radii given by $\{0,t,2t, 3t, \dots, Ct \}$, where $C$ and $t$ are design parameters (to be obtained later). The radius of the $m^{\text{th}}$ circle is given as $mr$. Note that the circle with radius zero corresponds to the origin. The points over each circle are chosen in an equidistance manner and the distance between any two points is chosen to be greater than or equal to $t$. Based on the aforementioned properties, in the following, we obtain the number and the radii of the circles.
\begin{itemize}
\item \textit{Number of circles $C+1$}:
Let $M_{m}$ be the maximum number of points, which can be placed over the $m^{\text{th}}$ circle. By convention we have $M_0 = 1$. It can be easily verified that for $M_m$ points on the $m^{\text{th}}$ circle to be at least with distance $t$ from each other, $M_m$ must satisfy
\begin{equation}\label{Eq_8}
   M_m = \left \lfloor \frac{\pi}{\arcsin \frac{1}{2m}} \right \rfloor.
\end{equation}
Given a message set of size $M$, we choose $C$ as the minimum value which encompasses $M$ points over the $C+1$ circumferences with radii $\{0,t,2t, ...,Ct \}$ under the constraint that the minimum distance between any two points is at least $t$. This is obtained by (thoroughly) filling the first $C$ circles with radii $\{0,t,2t, \ldots,(C-1)t \}$ according to (\ref{Eq_8}), and placing the remaining points (denoted by ${\bar{M}}_{C} $) in the circle with radius $Ct$. Therefore, the value of $C$ must satisfy the following equality
\begin{equation}
    M= 1 + \sum_{m=1}^{C-1}{{ M_m}} + {\bar{M}}_{C}  = 1 + \sum_{m=1}^{C-1}{{\left \lfloor \frac{\pi}{\arcsin \frac{1}{2m}} \right \rfloor + {\bar{M}}_{C}   }}.
\end{equation}
Note that $ {\bar{M}}_{C} =M - \sum_{m=0}^{C-1}{{ M_m}} \leq  {M}_{C}$.

\item \textit{The radii of circles}:
The parameter $t$ is chosen such that the average power of the constellation is equal to $P_a$. Accordingly, the parameter $t$ can be obtained as
%\begin{equation}
%      r^2 \cdot \frac{ \sum_{m=1}^{C-1} {\left \lfloor \frac{\pi}{\arcsin \frac{1}{2m}} \right \rfloor m^2  } + \tilde{N}_{\mathrm{C}} \cdot C^2}{M}.
%\end{equation}
%Hence, given a transmission power constraint $P_a$, the value of $r$ is given by
\begin{equation} \label{eq:circular}
    t = \sqrt{\frac{M P_a}{{\bar{M}_{\mathrm{C}} C^2 +\sum_{m=0}^{C-1} {\left \lfloor \frac{\pi}{\arcsin \frac{1}{2m}} \right \rfloor m^2  }}}}.
\end{equation}
%\item \textit{SER analysis}: In the high SNR regime, the SER of a constellation, denoted as $P_e$, can be approximated by
%\begin{equation}
%    P_{\mathrm{e}} =M_{\mathrm{min}} \mathcal{Q} \left(\frac{t}{\sqrt{2N_{0}}} \right),
%\end{equation}
%where $\mathcal{Q}(x)$ is the Q-function and $M_{\mathrm{min}}$ is the average number of points at distance $t$. For the considered constellation, a simple calculation verifies that $M_{\mathrm{min}} \leq 3$. It follows that, the SER reads as
%\begin{equation}
%    P_e \leq 3 \mathcal{Q} \left(\frac{t}{\sqrt{2N_0}} \right) = 3 \mathcal{Q} \left(  \sqrt{\frac{M \text{SNR}}{{\sum_{m=1}^{C-1} {\left \lfloor \frac{\pi}{\arcsin \frac{1}{2m}} \right \rfloor m^2  } + \bar{M}_{\mathrm{C}} C^2 }}  }\right).
%\end{equation}
\end{itemize}

\begin{figure}
\begin{centering}
\includegraphics[scale=0.4]{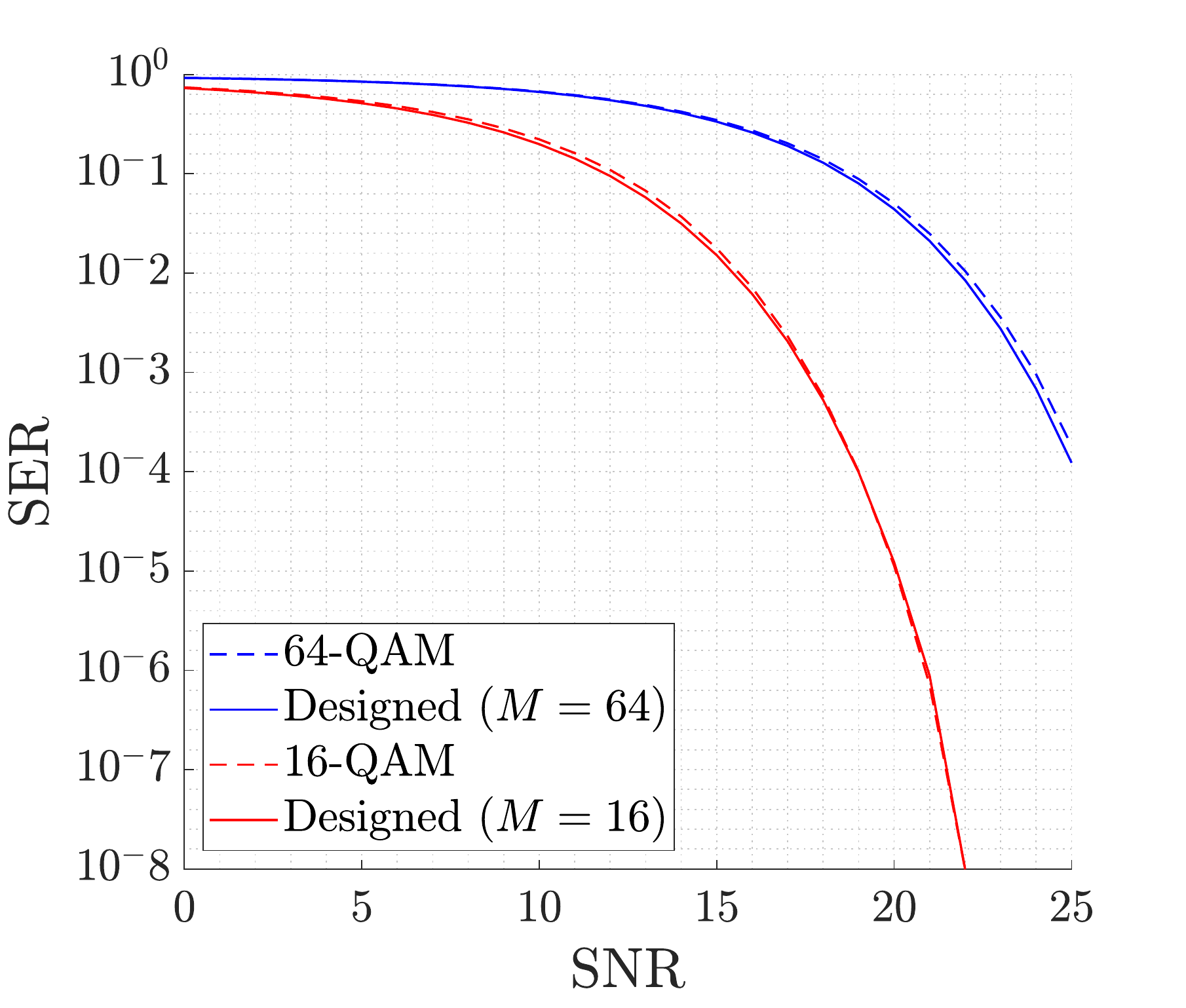}\vspace{-3mm}
\caption{SER vs. SNR between conventional QAM and the designed modulations for $M=16,~64$.}\label{Fig_Information_SER_vs_SNR}
\par\end{centering}
\vspace{-6mm}
\end{figure}

In Figure \ref{Fig_Information_SER_vs_SNR} the performance of the proposed information modulation (in terms of SER vs. SNR) is compared with the conventional QAM modulation for two different message set sizes, namely $M=16, ~64$. As it is illustrated, the performance of the designed modulations are very close to their QAM counterparts (with a slight improvement).

\subsubsection{PP-SWIPT Modulation}\label{Sec:SWIPT_Modulation}
The constellation explained in Section \ref{Sec:Information_Only_Modulation} are modified, such that some of the points of the external circles are allowed to have higher amplitudes and different phases. As a result of the increase in the amplitude for the complex points on the external circles, the remaining points of the constellation are to have lower amplitudes in order to satisfy the average power constraint.

The number of the complex points that are allowed to have higher amplitudes is denoted as $M_{on}$, and is obtained as a function of the probability $p_{on}^{\star}$ defined in (\ref{Eq_13}). Recall that $p_{on}^{\star}$ is the probability for the ON signal in On-Off keying signalling, which results in the maximum delivery power for the input of the form (\ref{Eq_9}). Given the value of the probability $p_{on}^{\star}$, the value of $M_{on}$ is obtained as
\begin{equation}\label{Eq_10}
      M_{on}=\argmin_{m\in\{1,\ldots,M\}} \left|p_{on}^{\star}-\frac{m}{M}\right|.
\end{equation}
The reason to obtain the parameter $M_{on}$ as in (\ref{Eq_10}) is justified by noticing that when the receiver power demand is at its maximum, the transmitter can resemble the optimal On-Off keying signalling by mapping all the $M_{on}$ complex points to the amplitude $\pmb{r}=|\pmb{x}|=\sqrt{MP_a/M_{on}}$ (distributed in an equidistance manner over a circle of radius $\sqrt{MP_a/M_{on}}$) and the remaining points to zero amplitude.

To choose the $M_{on}$ complex points, the complex symbols of the constellation are sorted amplitude-wise in a descending order. The first $M_{on}$ points are chosen and sorted in an ascending order with respect to their phase. The ordered complex points are denoted by $\{c_1,\ldots,c_{M_{on}}\}$. For each symbol $c_i,~i=1,\ldots,M_{on}$, we introduce two parameters, namely amplitude and phase shifts (denoted by $\alpha_i$ and $\beta_i$, respectively). The parameters $\alpha_i$ and $\beta_i$ are used in order to adjust the modulation according to the receiver information-power demands. Each one of the symbols $c_i,~i=1,\ldots,M_{on}$ is perturbed (denoted by $c_{i}^p$) as
\begin{equation}
      c_i^p=|c_i+\alpha_i|e^{\angle c_i+\beta_i},
\end{equation}
where the parameters $\alpha_i$ and $\beta_i$ are given as
\begin{subequations}\label{Eq_14}
\begin{align}
  \alpha_i&=\rho \left(\sqrt{\frac{MP_a}{M_{on}}}-|c_i|\right),\\
  \beta_i&=\rho\left(2(i-1)\pi/M_{on}-\angle c_i\right),
\end{align}
\end{subequations}
for $i=1,\ldots,M_{on}$ and $\rho\in[0,1]$. Note that $\rho$ is a parameter to control the receiver information-power demand (similar to $\lambda$ in (\ref{Eq_4})). That is, for $\rho=0$ we have the information modulation modelled in Section \ref{Sec:Information_Only_Modulation} and for $\rho =1$, we get $M-M_{on}$ zero symbols and $M_{on}$ symbols distributed in an equidistance manner over a circle with radius $\sqrt{MP_a/M_{on}}$.

\begin{figure}
\begin{centering}
\includegraphics[scale=0.26]{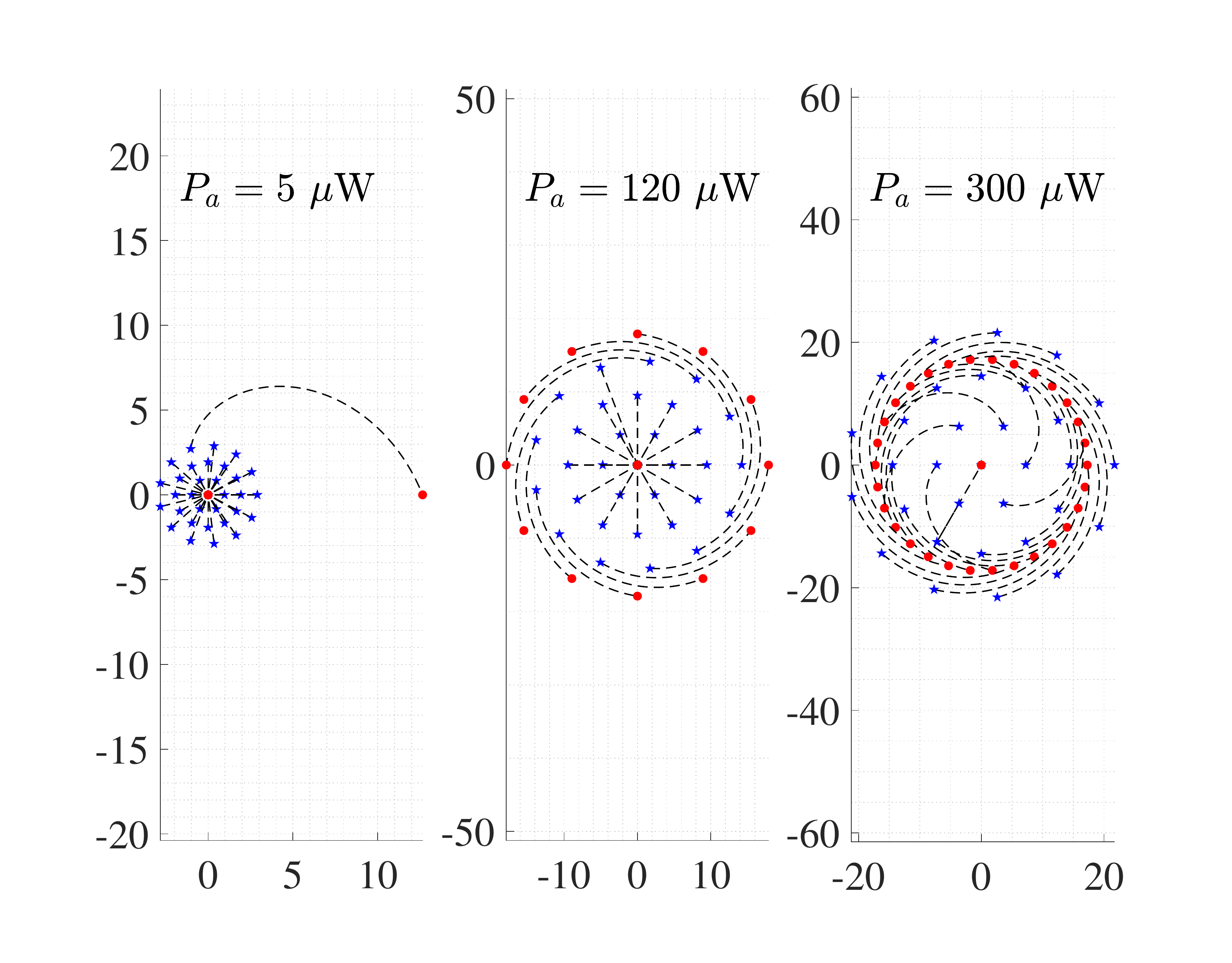}\vspace{-3mm}
\caption{Evolution of the designed modulations from information modulation (blue stars) to power modulation (red dots) for different average power constraints.}\label{Fig_Modulation_Evolution}
\par\end{centering}
\vspace{-6mm}
\end{figure}

In Figure \ref{Fig_Modulation_Evolution} the evolution of the information modulations to PP-SWIPT modulations is illustrated under different average power constraints, namely $P_a=5,~120,~300~\mu$W. For each plot, the blue stars correspond to the information modulation and the red dots correspond to the modulation when the receiver power demand reaches its maximum. The dashed line illustrates the locus of each symbol when the parameter $\rho$ varies from $0$ to $1$. The design of the modulation is such that the number of symbols that move away from zero increases with the average power constraint (due to (\ref{Eq_13}) and (\ref{Eq_10})), and ultimately when the receiver power demand is at its maximum, the transmit modulation is in the form of On-Off keying signalling of the form (\ref{Eq_9}).

\vspace{-1mm}
\subsection{Performance}\label{Sec:Modulation_Design_Performance}
\begin{figure}[!tbp]
  \centering
  \begin{subfigure}[b]{0.47\textwidth}
    \includegraphics[width=\textwidth]{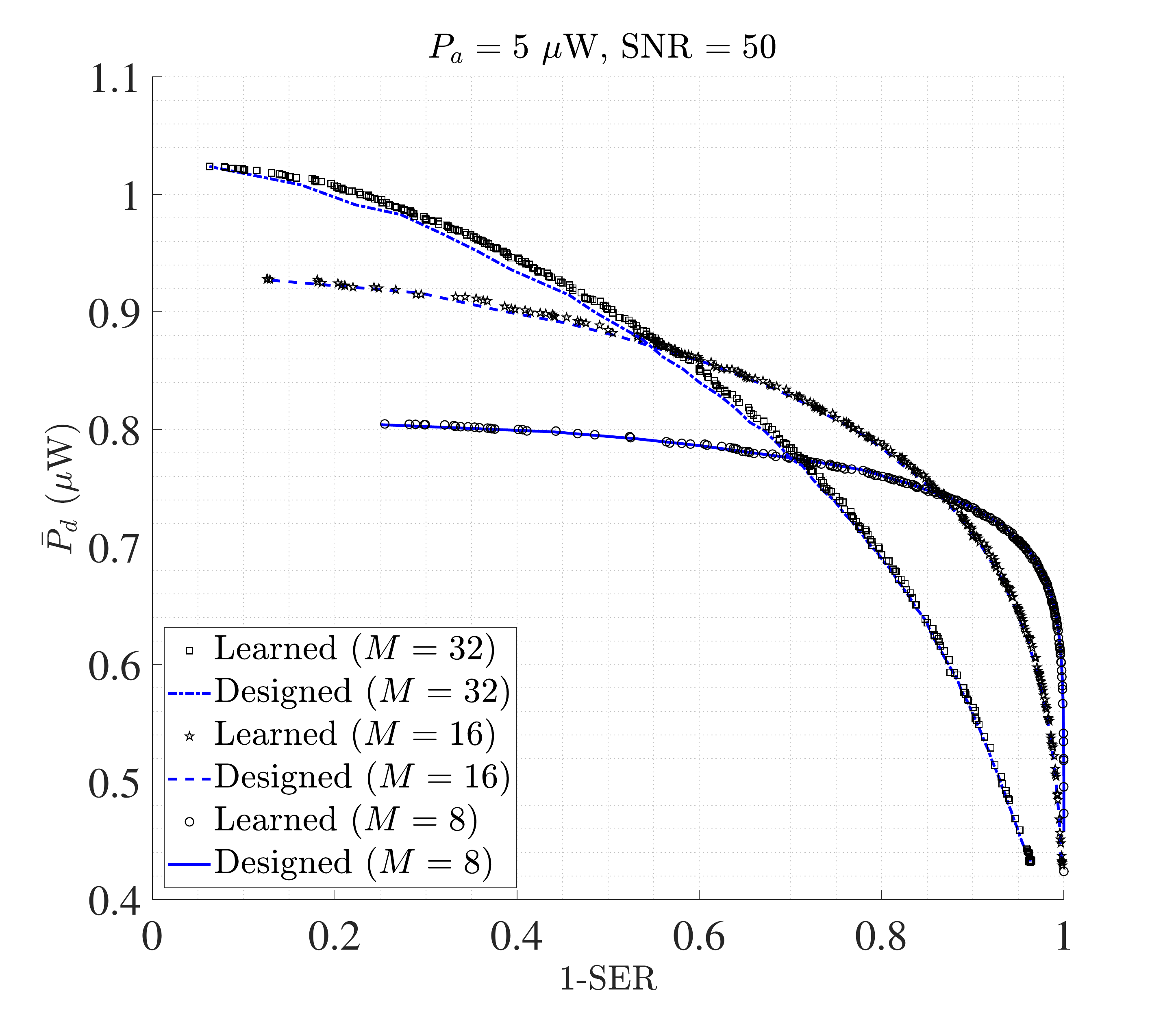}
    \caption{$P_a=5~\mu$W}\label{Fig_Pd_vs_SNR_0p005}
  \end{subfigure}
  \hfill
  \begin{subfigure}[b]{0.47\textwidth}
    \includegraphics[width=\textwidth]{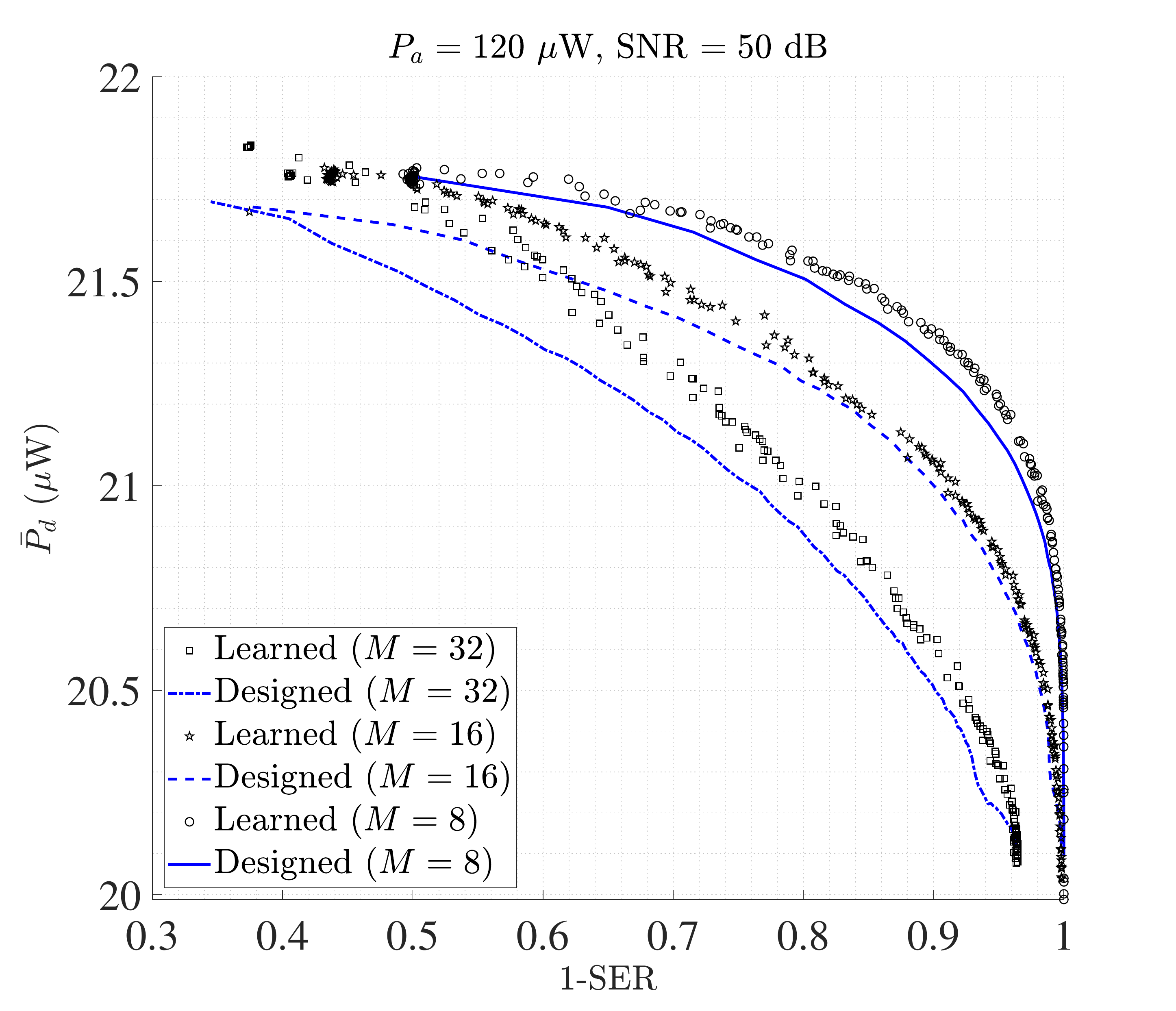}
    \caption{$P_a=120~\mu$W}\label{Fig_Pd_vs_SNR_0p1}
  \end{subfigure}
  \caption{Delivered power $\bar{P}_d$ versus complementary symbol error rate $1-\text{SER}$ tradeoff for message sizes $M=8,~16$ and $32$ with a) $P_a=5~\mu$W and b) $P_a=120~\mu$W. Black dots and blue lines represent the performance corresponding to learning and algorithmic approach, respectively.}
  \vspace{-6mm}
\end{figure}

In Figures \ref{Fig_Pd_vs_SNR_0p005} and \ref{Fig_Pd_vs_SNR_0p1} the delivered power $\bar{P_d}$ versus complementary symbol error rate ($1-\text{SER}$) for different message set sizes, namely $M=8$, $16$ and $32$, with SNR$=50$ ($16.98$ dB) are illustrated under average power constraints $P_a=5~\mu$W and $P_a=120~\mu$W, respectively. In each plot, the blue (solid and dashed) lines and the black (circle, star and square) dots illustrate the performance corresponding to learning (in Section \ref{Sec:Modulation_Design_Learning}) and algorithmic (in Section \ref{Sec:Modulation_Design_Algorithmic}) approaches, respectively. The performance of the designed modulations is very close to their learned counterparts. Additionally, the designed modulations do not require training, and accordingly, represent high adaptivity to change of system design parameters.

%\begin{figure}
%\begin{centering}
%\includegraphics[scale=0.25]{PP_Pd_vs_SNRc_0p005.pdf}
%\caption{Illustration of the tradeoff between the  average delivered power $\bar{P}_d$ and complementary symbol error rate $1-\text{SER}$ for message size $M=8,~16$ and $32$ under average power constraint $P_a=5~\mu$W. Black (squares, stars and circles) dots and blue (solid and dashed) lines represent the performance corresponding to learning and algorithmic approach, respectively. }\label{Fig_Pd_vs_SNR_0p005}
%\par\end{centering}
%\vspace{0mm}
%\end{figure}
%
%\begin{figure}
%\begin{centering}
%\includegraphics[scale=0.25]{PP_Pd_vs_SNRc_0p1.pdf}
%\caption{Illustration of the tradeoff between the  average delivered power $\bar{P}_d$ and complementary symbol error rate $1-\text{SER}$ for message size $M=8,~16$ and $32$ under average power constraint $P_a=120~\mu$W. Black (squares, stars and circles) dots and blue (solid and dashed) lines represent the performance corresponding to learning and algorithmic approach, respectively.}\label{Fig_Pd_vs_SNR_0p1}
%\par\end{centering}
%\vspace{0mm}
%\end{figure}

%{\color{red}how to send information with lower SER with the power modulation}
\vspace{-1mm}
\begin{rem}\label{rem:1}
One main downside of the PP-SWIPT modulations (obtained via both learning and algorithmic approaches) is that they fail to transmit information for high receiver power demands. This issue can be partially alleviated by allowing the transmitter to send blocks of symbols per message. Assume the receiver power demand is at its maximum (requiring the transmitter to follow a signalling of the form (\ref{Eq_9}) with optimal probability $p_{on}^{\star}$ in (\ref{Eq_13})). Rather than transmitting one symbol per message (with rate $\log_2 M$), blocks of length $n>1$ are transmitted. In each block, we have either a zero or a nonzero symbol. The number of nonzero symbols (denoted as $N_{on}$) in a block of length $n$ and the value (denoted as $r_{on}$) of nonzero symbol is determined as
\begin{align}
N_{on} &= \argmin_{l\in\{1,\ldots,n\}} \left|p_{on}^{\star}-\frac{l}{n}\right|,\\
r_{on}&= \sqrt{n P_a/N_{on}}.
\end{align}
As $n$ grows large, we have a better approximation of the probability $p_{on}^{\star}$ in each block. Additionally, since in each block there are $N_{on}$ nonzero symbols, we can have $M\leq {n \choose N_{on}}$ distinguishable blocks of length $n$ (equivalent to the maximum number of messages to be transmitted). For the information receiver, the message estimation is simplified to detecting the position of nonzero symbols in the received signal. The size of the message set $M$ and the length of the transmitted block $n$ are parameters that are to be determined based on the desired SER, the delay tolerance and complexity of the receiver. The design of such codes (on its own) is an interesting topic of research and we postpone investigating them further in future works.
\end{rem}

\vspace{-1mm}
\section{Modulation Design for Multi-User SWIPT}\label{Sec:Extension to Multi-user SWIPT}
In this section, we generalize the learning approach to modulation design for multi-user SWIPT scenarios. In particular, we consider two user degraded \textit{Broadcast Channel} (BC), \textit{Multiple Access Channel} (MAC), and \textit{Interference Channel} (IC)\footnote{Similar to PP-SWIPT modulation design, for the different system models studied in this section, we have implemented both fully connected NN as well as convolutional NN structures. It is observed that the final results are similar, with the main difference that implementing fully connected NN take relatively longer time to converge to the final solution. Since there are detailed discussions in the literature on how to implement communications systems as NN-based autoencoders, for brevity, we do not discuss about the details of the implemented structures.}.

\vspace{-1mm}
\subsection{Degraded Broadcast Channel (BC)}\label{Sec_BC}
\subsubsection{Transmitter}
The transmitter maps $K$ independent messages\footnote{For brevity, we do not assume common messages.} $\{\pmb{s}_j\}_{j=1}^K$ (each of them chosen out of an alphabet $\mathcal{M}_j = \{1,2, \cdots, M_j \}$) into a codeword $\pmb{x}^n \in \mathcal{X}^{n},~\mathcal{X}^n\in \mathbb{C}^n$ through a mapping function denoted as $g_{\Theta_{\mathrm{T}}}(\cdot)  : \mathcal{M}_1 \times \mathcal{M}_2 \times \dots \times \mathcal{M}_{K} \rightarrow \mathbb{C}^n $. In general, the alphabet sets are of different sizes for different receivers.

In the implementation of the transmitter, the input layer is obtained by concatenating into a vector the $K$ binary vectors (denoted as $\pmb{s_{b}}_j,~j=1,\ldots,K$) corresponding to the binary representations of $K$ messages to be transmitted to the $K$ receivers. Therefore, the input layer consists of $\sum_{j=1}^K{\lceil\log_2 M_j\rceil}$ units (in this paper, we assume $M_j$ as a power of two for $j =1,2,\cdots,K$).

\subsubsection{Receivers}
The $K$ receivers are indexed as $\textit{Rx}_{j},~j =1,2,\cdots,K$. Each receiver ${\textit{Rx}}_j$ estimates its corresponding message by mapping the received signal $\pmb{y}_j^n$ into an $M_j$-dimensional output probability vector $\pmb{\hat{s}_o}_{j}$ using the information receiver mappings denoted as $h_{j,\Theta_{\textit{R},j}}(\cdot)  : \mathbb{C}^n \rightarrow \mathbb{R}^{M_j} $ for $j=1,\ldots, K$, and harvests the energy of its received signal $\pmb{y}_j^n$.

\subsubsection{Objective function}
The probabilistic objective function is defined as\footnote{Note that in the degraded BC, the average error probability can be upper bounded by the summation of the average error probability of each receiver.}
\begin{align}\label{Eq_1}
L(\Theta_T,\Theta_{R_1},\ldots,\Theta_{R_K}) = \sum\limits_{j=1}^{K}\mathbb{E}\left[-\log{p_{\hat{\pmb{s}}_j|\pmb{s}_j}}(\hat{\pmb{s}}_j=\pmb{s}_j|\pmb{s}_j)+\frac{\lambda}{P_{d,j}(\pmb{s}_1,\ldots,\pmb{s}_K)}\right],
\end{align}
where the expectation is over the randomness of the transmitted messages and the channel noise, and $P_{d,j}(\pmb{s}_1,\ldots,\pmb{s}_K)$ is the harvested power at the $j^{\text{th}}$ receiver $\textit{Rx}_j$. The objective function in (\ref{Eq_1}) is approximated with $L_a(\Theta_T,\Theta_{R_1},\ldots,\Theta_{R_K})$ defined as\footnote{Defining a cost function is potentially dependent on many elements, such as the type of the receiver (whether it is an energy harvester or an information receiver or both), the requirement of different receivers (for instance one requesting for energy and the other requesting information) and so on. In this work, we relate the power demands of the receivers to one parameter, namely $\lambda$. This is mainly due to increasing the representability of the results.}
\begin{align} \label{cost_function_broadcast_channel}
L_a(\Theta_T,\Theta_{R_1},\ldots,\Theta_{R_K}) = \frac{1}{|\mathcal{B}|} \sum_{l\in \mathcal{B}} \sum_{j=1}^{K} \mathcal{L}(s_{o,j}^{(l)}, \hat{s}_{o,j}^{(l)}) +\frac{\lambda}{P_{d,j}(s_1^{(l)},\ldots,s_K^{(l)})},
\end{align}
where $\mathcal{L}(s_{o,j}^{(l)}, \hat{s}_{o,j}^{(l)}) = -\sum_{i=1}^M {s_{o,j,i}^{(l)}  \log \hat{s}_{o,j,i}^{(l)}}$ is the cross entropy function modelling the information loss at the $j^{\text{th}}$ receiver $\textit{Rx}_j$, $s_{o,j,i}^{(l)}$ and $\hat{s}_{o,j,i}^{(l)}$ indicate the $i^{\text{th}}$ entry of the vectors $s_{o,j}^{(l)}$ and $\hat{s}_{o,j}^{(l)}$, respectively. $\mathcal{B}$ is the randomly drawn training data and the superscript $(l)$ indicates the $l^{\text{th}}$ training sample. Note that each training sample is a vector of length $2K$ consisting of $K$ transmit messages (randomly realized from the sets $\mathcal{M}_j$ for $j=1,2, \cdots, K$) and their corresponding true estimations at the output of the $K$ receivers.

\vspace{-1mm}

\subsection{Multiple Access Channel (MAC)} \label{Multiple_Access_Channel}
\subsubsection{Transmitters}
$K$ transmitters, denoted as $\textit{Tx}_{j},~j =1,2,\cdots,K$, communicate independent messages to a receiver. Chosen out of an alphabet of $M_{j}$ symbols $\mathcal{M}_j = \{1,2, \cdots, M_j \}$, each transmitter $\textit{Tx}_{j}$ transmits a message $\pmb{s}_j$ (by first converting it into a binary vector $\pmb{s_b}_j$),  by mapping it into a codeword $\pmb{x}_j^n$ through a mapping function denoted as  $g_{\Theta_{\textit{T},j}}(\cdot) : \mathcal{M}_j \rightarrow \mathbb{C}^n,~j=1,\ldots,K$, where $\Theta_{\textit{T},j}$ represents the set of parameters used for characterizing the structure of the NN at $\textit{Tx}_j$.

\subsubsection{Receiver}
The receiver maps the received codeword $\pmb{y}^n$ into a vector with $\sum_{j=1}^K{M_j}$ entries. This vector is then split into $K$ output probability vectors $\pmb{\hat{s}_o}_j$ (each with length $M_j$) corresponding to each transmitted message $\pmb{s}_j,~~j=1,\ldots,K$. In addition to the information receiver, the EH harvests the energy corresponding to the received signal $\pmb{y}^n$.

\subsubsection{Objective Function}
The probabilistic objective function is defined similarly to (\ref{Eq_1}). The approximated objective function $L_a(\Theta_{T,1},\ldots,\Theta_{T,K},\Theta_R)$ is then defined as
\begin{align} \label{cost_multiple_access_channel}
L_a(\Theta_{T,1},\ldots,\Theta_{T,K},\Theta_R) = \frac{1}{|\mathcal{B}|}\sum_{l\in \mathcal{B}}  \sum_{j=1}^{K}\mathcal{L}(s_{o,j}^{(l)}, {\hat{s}}_{o,j}^{(l)}) +\frac{\lambda}{P_{d,j}(s_1^{(l)},\ldots,s_K^{(l)})}.
\end{align}

\vspace{-1mm}
\subsection{Interference Channel (IC)} \label{Introduction}

\subsubsection{Transmitters}
$K$ transmitters, denoted as $\textit{Tx}_{j},~j =1,2,\cdots,K$, communicate independent messages $\pmb{s}_j\in\mathcal{M}_j=\{1,\ldots,M_j\},~j=1,\ldots,K$ to $K$ receivers, denoted as $\textit{Rx}_{j},~j =1,2,\cdots,K$. Each transmitter $\textit{Tx}_{j}$ maps a message $\pmb{s}_j$ (by first converting it into a binary vector $\pmb{s_b}_j$) into a codeword $\pmb{x}_j^n$ via a mapping function denoted as  $g_{\Theta_{\textit{T},j}}(\cdot) : \mathcal{M}_j \rightarrow \mathbb{C}^n,~j=1,\ldots,K$, where $\Theta_{\textit{T},j}$ represents the set of parameters used for characterizing the structure of the $j^{\text{th}}$ transmitter $\textit{Tx}_j$.

\subsubsection{Receiver}
Receiver ${\textit{Rx}}_j$, receives the noisy signal $\pmb{y}_j^n=\sum_{d=1}^{K}a_{d,j}\pmb{x}_d^n+\pmb{w}_j^n$, where $a_{d,j}$ is the gain of the channel from transmitter $\textit{Tx}_d$ to $\textit{Rx}_j$ with $a_{j,j}=1,~j=1,2,\cdots,K$. Each receiver ${\textit{Rx}}_j$ maps the received signal $\pmb{y}_j^n$ into an $M_j$-dimensional probability vector $\hat{\pmb{s}}_{o,j}$ via the receiver mapping function denoted as  $h_{j,\Theta_{\textit{R},j}}(\cdot)  : \mathbb{C}^n \rightarrow \mathbb{R}^{M_j} $ for $j=1,\ldots, K$, and harvests the energy corresponding to $\pmb{y}_j^n$.

\subsubsection{Cost function}
The probabilistic objective function is defined similarly to (\ref{Eq_1}). The approximated objective function $L_a(\Theta_{T,1},\ldots,\Theta_{T,K},\Theta_{R,1},\ldots,\Theta_{R,K}) $ is then defined as
\begin{align} \label{cost_multiple_access_channel}
L_a(\Theta_{T,1},\ldots,\Theta_{T,K},\Theta_{R,1},\ldots,\Theta_{R,K}) = \frac{1}{|\mathcal{B}|} \sum_{l\in \mathcal{B}} \sum_{j=1}^{K}\mathcal{L}(s_{o,j}^{(l)}, {\hat{s}}_{o,j}^{(l)}) +\frac{\lambda}{P_{d,j}(s_1^{(l)},\ldots,s_K^{(l)})}.
\end{align}

\vspace{-1mm}

\subsection{Numerical Results} \label{Sex:Multiuser_Modulation_learned_results}
In the following, the multi-user SWIPT modulations obtained via learning approach for the two user ($K=2$) BC, MAC, IC are illustrated. The learning approach for designing SWIPT modulations does not seem a feasible approach in practice. This is mainly due to the fact that by increasing the number of users as well as message set sizes, the training time increases exponentially. Accordingly, if the receiver information-power demand changes faster than the training time, the resulting modulation is of no use. However, the learned results can be utilized to extract some properties of the solutions, which can be used for designing algorithmic modulations (similar to Section \ref{Sec:Modulation_Design_Algorithmic}).

\subsubsection{BC}
\begin{figure}[!tbp]
  \centering
  \begin{subfigure}[b]{0.495\textwidth}
    \includegraphics[width=\textwidth]{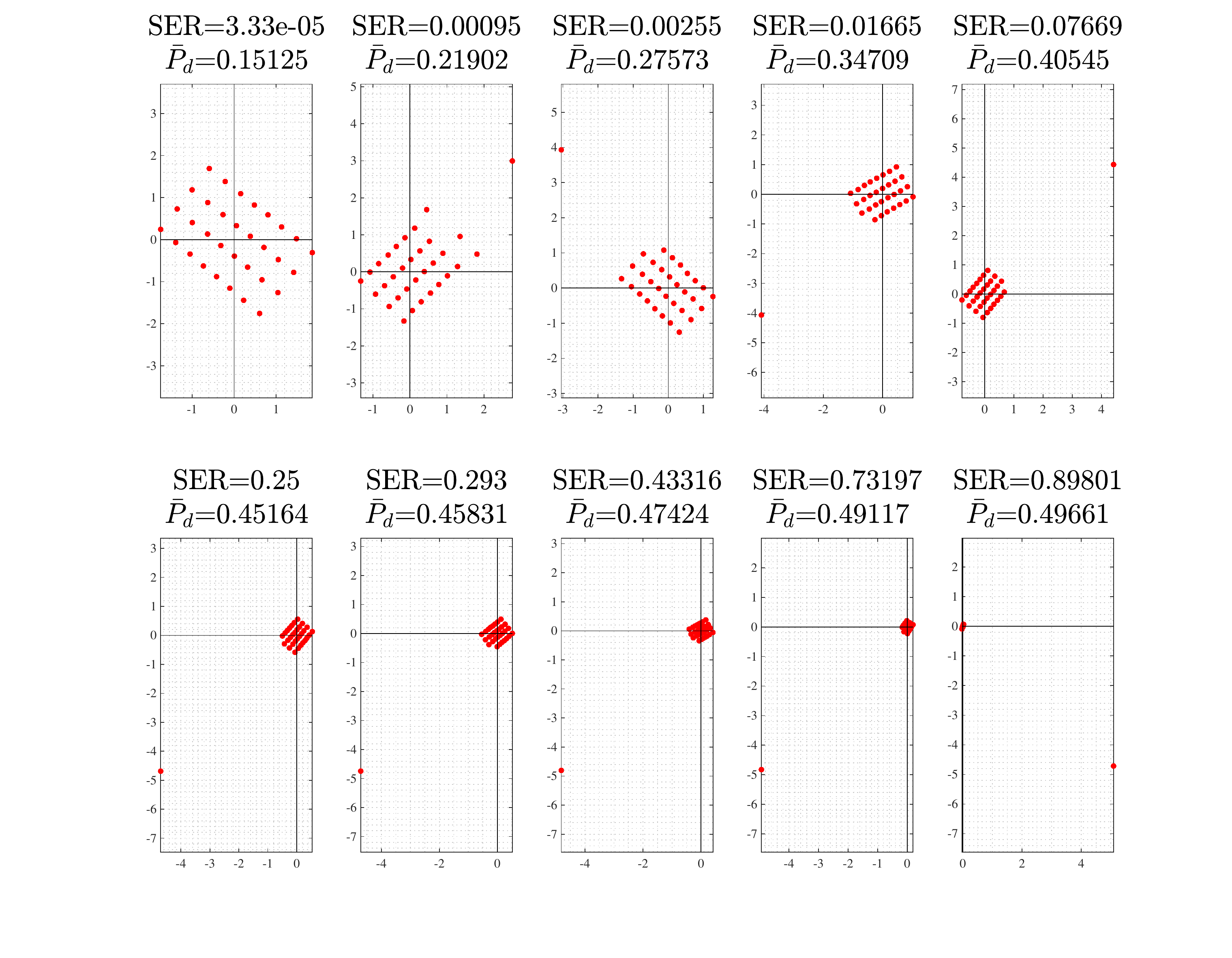}
    \caption{$P_a=5~\mu$W}\label{BC_0015}
  \end{subfigure}
  \hfill
  \begin{subfigure}[b]{0.495\textwidth}
    \includegraphics[width=\textwidth]{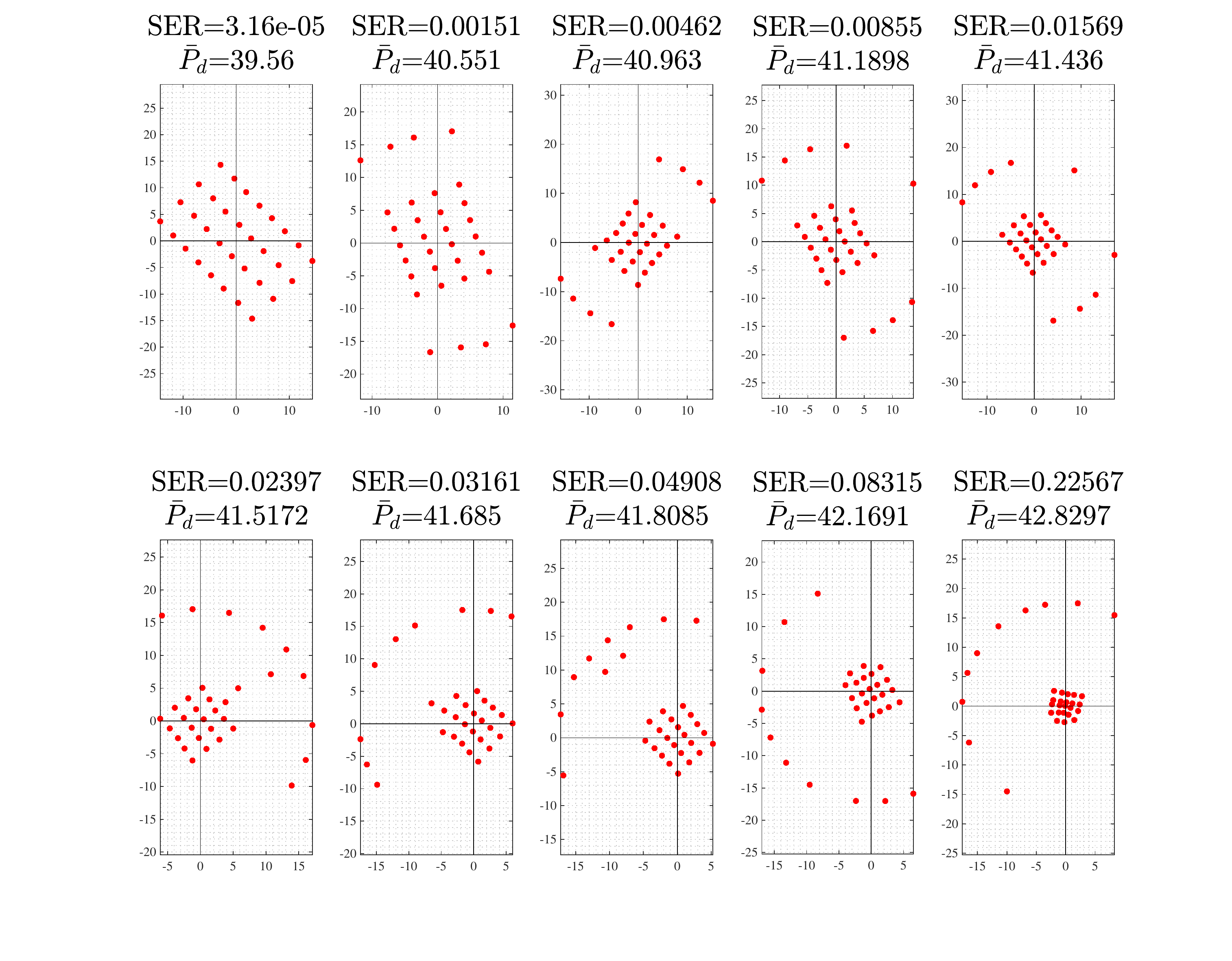}
    \caption{$P_a=100~\mu$W}\label{BC_01}
  \end{subfigure}
  \caption{Learned modulations for degraded BC under an average power constraint a) $P_a=5~\mu$W b) $P_a=100~\mu$W and with message set sizes $M_1=8,~M_2=4$ and $\text{SNR}_1=100$, $\text{SNR}_2=50$.}
  \vspace{-6mm}
\end{figure}

In Figures \ref{BC_0015} and \ref{BC_01}, the learned modulations for degraded BC are illustrated under an average power constraint $P_a=1.5~\mu$W and $P_a=100~\mu$W, respectively. In both Figures, we have $M_1=8$ (strong receiver) and $M_2=4$ (weak receiver). When the receiver demand is merely information (the first plot in Figures \ref{BC_0015} and \ref{BC_01}) the transmit modulation is a rectangular constellation resembling a superposition coding \cite{elGamal:book} (the strong receiver estimates both messages, whereas the weak receiver estimates only the row in which its corresponding message is transmitted). We note that the tilting angle of the rectangular constellation is arbitrary since it does not affect harvesting/decoding. As the power demand of the users increases, the modulation reforms in a way that some of the transmit symbols move away from the origin, where in the extreme scenario, the transmit modulation resembles an On-Off keying signalling (with some symbols collapsing on top of each other and the other symbols moving away from the origin and distributed equidistantly). As discussed in Section \ref{Sec:Learned_Model}, the number of symbols that move away from the origin (by increasing the receivers' power demand) is dependent on the input power of the EH (for example, under an average power constraint of $P_a=100~\mu$W from (\ref{Eq_9}) and (\ref{Eq_13}), it is verified that the number of symbols moving away from the origin is around $10$).

%\begin{figure}
%\begin{centering}
%\includegraphics[scale=0.3]{BC_nx1_8_nx2_4_Constel_0p0015.pdf}
%\caption{Learned modulations for degraded BC under an average power constraint $P_a=5~\mu$W and with message set sizes $M_1=8,~M_2=4$ and $\text{SNR}_1=100$, $\text{SNR}_2=50$.}\label{BC_0015}
%\par\end{centering}
%\vspace{0mm}
%\end{figure}

%\begin{figure}
%\begin{centering}
%\includegraphics[scale=0.3]{BC_nx1_8_nx2_4_Constel_0p1.pdf}
%\caption{Learned modulations for degraded BC under an average power constraint $P_a=100~\mu$W and with message set sizes $M_1=8,~M_2=4$ and $\text{SNR}_1=100$, $\text{SNR}_2=50$.}\label{BC_01}
%\par\end{centering}
%\vspace{0mm}
%\end{figure}

\subsubsection{MAC}
\begin{figure}[!tbp]
  \centering
  \begin{subfigure}[b]{0.49\textwidth}
    \includegraphics[width=\textwidth]{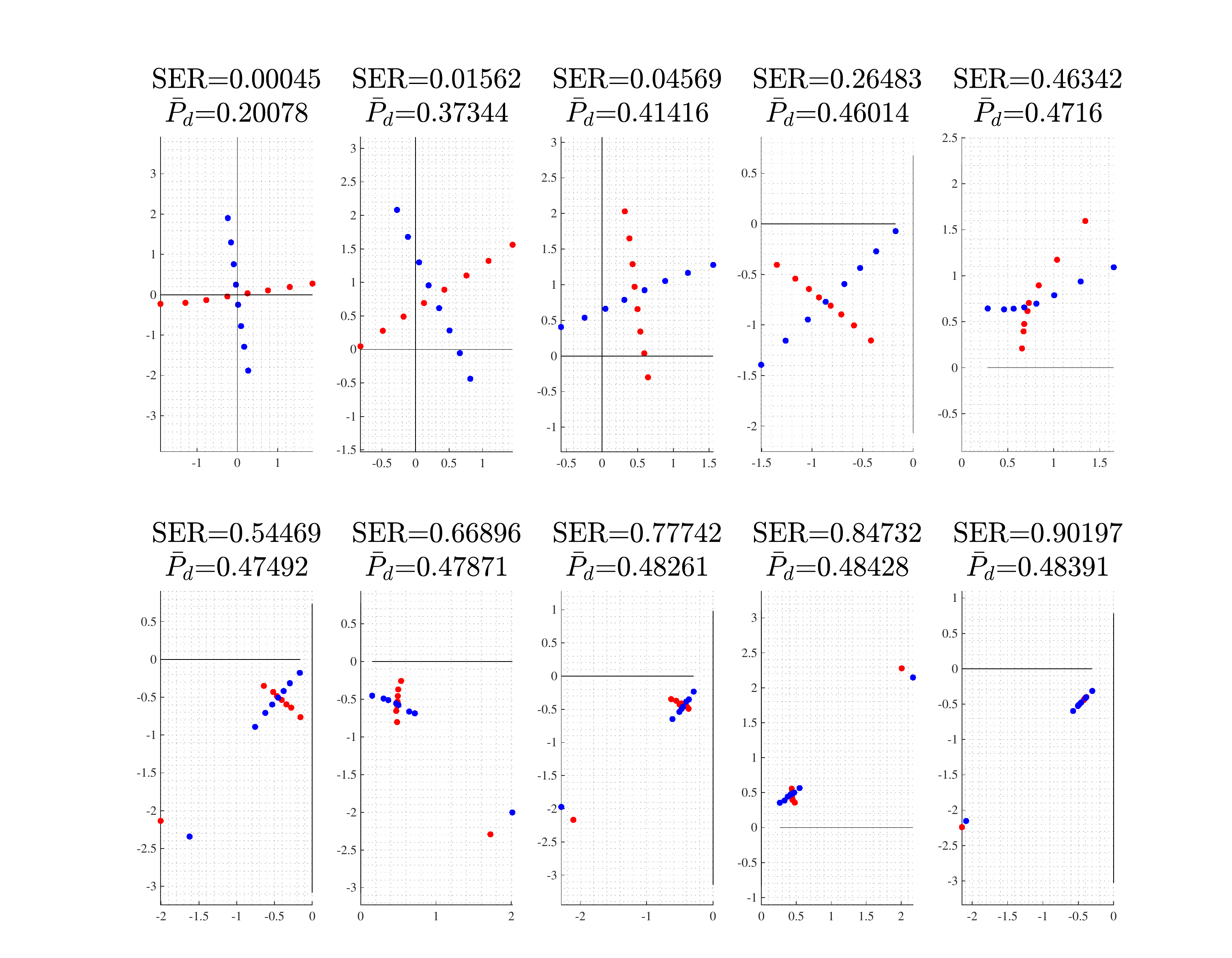}
    \caption{$P_a=1.5~\mu$W (per transmitter)}\label{MAC_0015}
  \end{subfigure}
  \hfill
  \begin{subfigure}[b]{0.49\textwidth}
    \includegraphics[width=\textwidth]{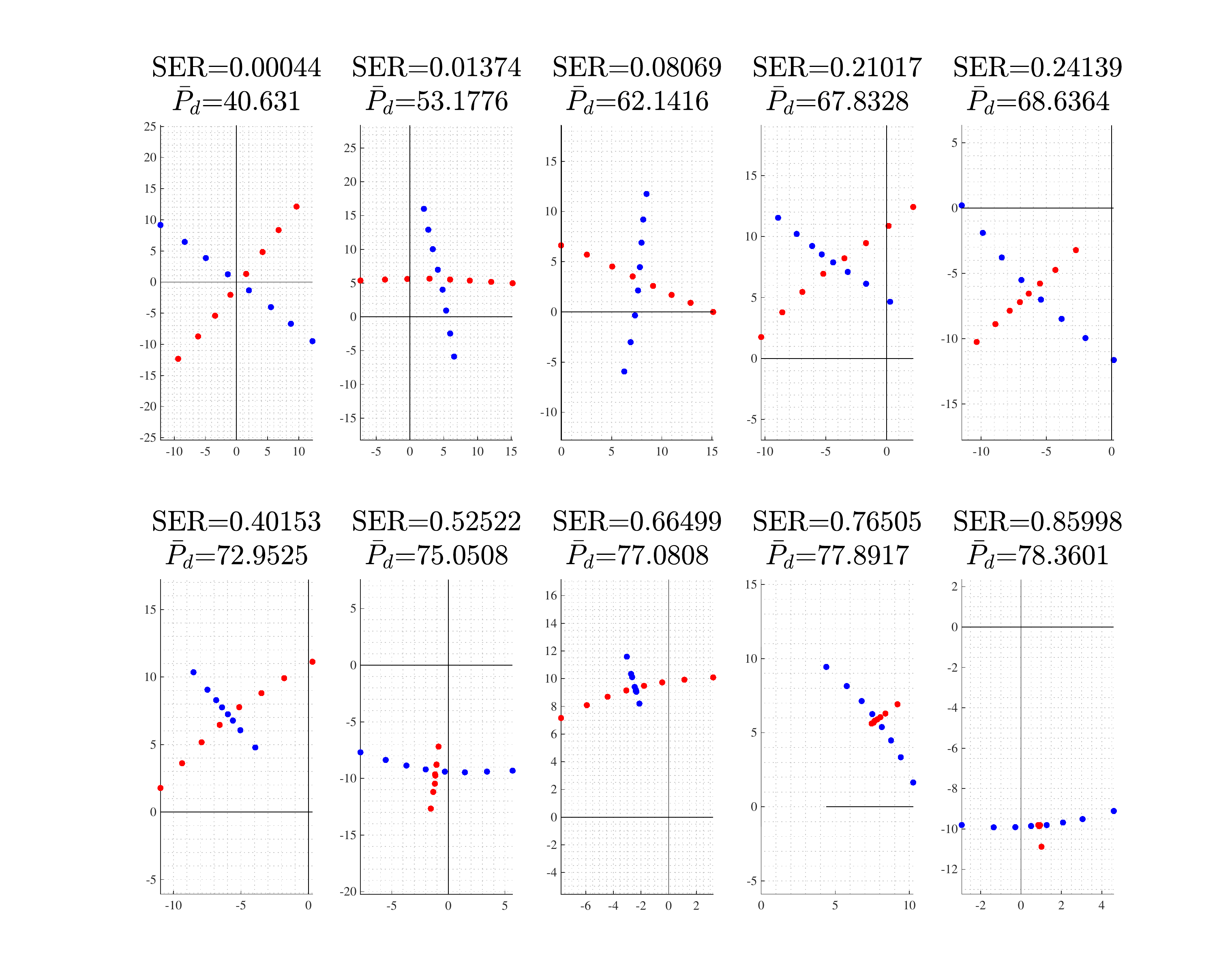}
    \caption{$P_a=100~\mu$W (per transmitter)}\label{MAC_01}
  \end{subfigure}
  \caption{Learned modulations for MAC under an average power constraint a) $P_a=1.5~\mu$W and b)$P_a=100~\mu$W and with message set sizes $M_1=8,~M_2=8$ and $\text{SNR}=50$ (per transmitter).}
  \vspace{-6mm}
\end{figure}

In Figures \ref{MAC_0015} and \ref{MAC_01}, the learned modulations for MAC are illustrated under an average power constraint $P_a=1.5~\mu$W and $P_a=100~\mu$W (per transmitter), respectively. In both Figures, we have $M_1=M_2=8$, and in each plot, the red and blue dots correspond to the transmit constellation of user $1$ and user $2$, respectively. Note that in both Figures, when the receiver demand is merely information, the channel input is a rectangular constellation (rotated cartesian product of the transmit constellation of the users). Indeed, the transmitters' constellations are formed in an orthogonal way so that estimating the transmit message of each user can be implemented by projecting the received signal on the axis corresponding to each transmitter's modulation. As the receiver power demand increases, for low average power constraints (see Figure \ref{MAC_0015} ), both of the transmitters contribute in forming an On-Off keying signalling by moving some of their transmit symbols away from the origin. However, for high average power constraints (see Figure \ref{MAC_01}), as the receiver power demand increases, the transmit modulations change as follows. First, the modulation symbols of both transmitters move away from the origin, second, one transmitter is merely transmitting power by amalgamating the symbols, while the other keep the transmit symbols as far apart as possible to reduce the ultimate SER.

\vspace{-1mm}
\begin{rem}
The MAC setup considered here is symmetric in terms of the design parameters, that is the channel additive noise variance and the transmitters' average power constraints. However, as it is illustrated in the Figures \ref{MAC_0015} and \ref{MAC_01}, the well-performing solution is not necessarily a symmetric one.
\end{rem}
\vspace{-1mm}

%\begin{figure}
%\begin{centering}
%\includegraphics[scale=0.3]{MAC_nx1_8_nx2_8_Constel_0p0015.pdf}
%\caption{Learned modulations for MAC under an average power constraint $P_a=1.5~\mu$W (per transmitter) and with message set sizes $M_1=8,~M_2=8$ and $\text{SNR}=50$ (per transmitter).}\label{MAC_0015}
%\par\end{centering}
%\vspace{0mm}
%\end{figure}
%
%\begin{figure}
%\begin{centering}
%\includegraphics[scale=0.3]{MAC_nx1_8_nx2_8_Constel_0p1.pdf}
%\caption{Learned modulations for MAC under an average power constraint $P_a=100~\mu$W (per transmitter) and with message set sizes $M_1=8,~M_2=8$ and $\text{SNR}=50$ (per transmitter).}\label{MAC_01}
%\par\end{centering}
%\vspace{0mm}
%\end{figure}

\subsubsection{IC}
\begin{figure}[!tbp]
  \centering
  \begin{subfigure}[b]{0.49\textwidth}
    \includegraphics[width=\textwidth]{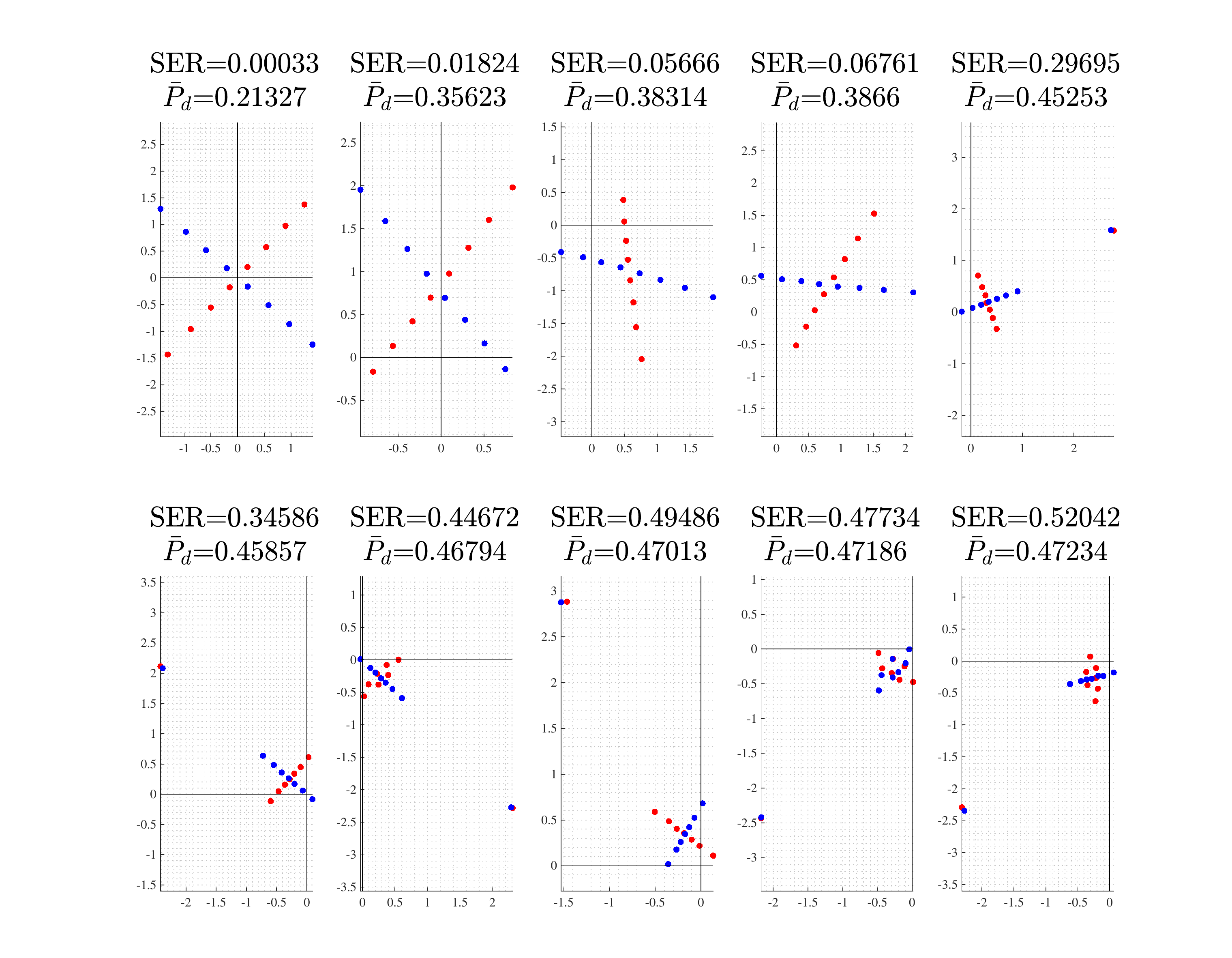}
    \caption{$P_a=1.5~\mu$W (per transmitter)}\label{IC_0015}
  \end{subfigure}
  \hfill
  \begin{subfigure}[b]{0.49\textwidth}
    \includegraphics[width=\textwidth]{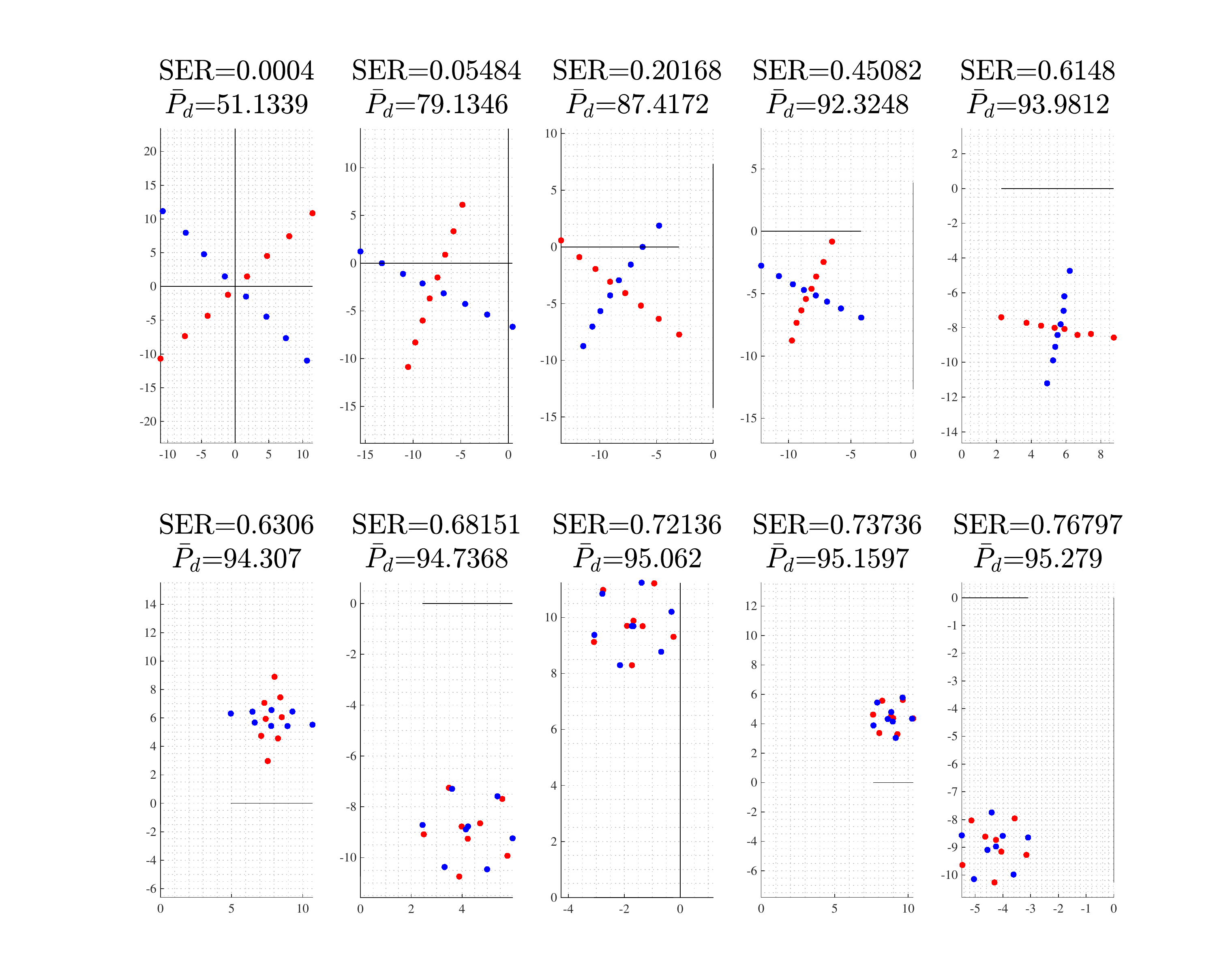}
    \caption{$P_a=100~\mu$W (per transmitter)}\label{IC_01}
  \end{subfigure}
  \caption{Learned modulations for IC under an average power constraint a) $P_a=1.5~\mu$W b) $P_a=100~\mu$W and with message set sizes $M_1=8,~M_2=8$, with SNR$=50$ (per link) and cross link gain $a_{12}=a_{21}=0.5$.}
  \vspace{-6mm}
\end{figure}

In Figures \ref{IC_0015} and \ref{IC_01}, the learned modulations for IC are illustrated under an average power constraint $P_a=1.5~\mu$W and $P_a=100~\mu$W (per transmitter), respectively. In both figures, we have $M_1=M_2=8$ and in each plot, the red and blue dots correspond to the transmit constellation of user $1$ and user $2$, respectively. Here, we consider the weak interference\footnote{The weak IC channel assumption is for the sake of representation. The results does not necessarily hold for other types of IC channels.} channel with two users where the gain in the cross link is less than one (here the gain is assumed $0.5$). The observations are mainly similar to the ones obtained for the MAC scenario. That is the higher the power demand, the greater the nonzero mean of the transmit modulation. Also, for dominant information demands at the receiver (the first four upper plots in Figures \ref{IC_0015} and \ref{IC_01}), the transmitters follow an orthogonal transmit modulation. One main difference compared to the MAC modulations is that, in the high channel input power regime ($P_a = 100~\mu$W), since we have two different receivers that demand for information as well as power, the transmitters do not modulate as the MAC scenario (one sending power and the other sending information). Accordingly, the transmitters favour a circular nonzero mean modulation so that some information is sent as well.

\vspace{-1mm}
\begin{rem}
The learned modulations for multi-user SWIPT can be used to better understand the signal design (even for scenarios where the receiver demands merely for information). Accordingly, an algorithmic approach similar to that presented in Section \ref{Sec:Modulation_Design_Algorithmic} can be utilized to design modulations that do not require training. In this paper we have not focused on algorithmic multi-user SWIPT modulation desing.
\end{rem}
\vspace{-1mm}

%\begin{figure}
%\begin{centering}
%\includegraphics[scale=0.3]{IC_nx1_8_nx2_8_Constel_0p0015.pdf}
%\caption{Learned modulations for IC under an average power constraint $P_a=1.5~\mu$W (per transmitter) and with message set sizes $M_1=8,~M_2=8$, with SNR$=50$ (per link) and cross ling gain $a_{12}=a_{21}=0.5$.}\label{IC_0015}
%\par\end{centering}
%\vspace{0mm}
%\end{figure}
%
%\begin{figure}
%\begin{centering}
%\includegraphics[scale=0.3]{IC_nx1_8_nx2_8_Constel_0p1.pdf}
%\caption{Learned modulations for IC under an average power constraint $P_a=100~\mu$W (per transmitter) and with message set sizes $M_1=8,~M_2=8$, with SNR$=50$ (per link) and cross ling gain $a_{12}=a_{21}=0.5$.}\label{IC_01}
%\par\end{centering}
%\vspace{0mm}
%\end{figure}

\vspace{-1mm}
\section{Coded Modulation for SWIPT}\label{Sec:Extension_to_coded_modulation}
The learned and designed constellations in Sections \ref{Sec:SWIPT_PP_Modulation_Design} demonstrate a strong drawback when the receiver energy demands dominate its information demands. As it is illustrated in Figures \ref{Fig_Pd_vs_SNR_0p005} and \ref{Fig_Pd_vs_SNR_0p1}, the SER of the transmitted information increases drastically with the receiver power demand. This is mainly due to the fact that by increasing the receiver power demand, some of the symbols of the transmitted constellation converge towards zero amplitude (see Figure \ref{Fig_Modulation_Evolution}), making the decoding process challenging. This issue can be partially overcome by considering longer lengths for the transmitted codewords $\pmb{x}^n$, i.e., $n>1$.

In the following, we extend the SWIPT modulation design in Section \ref{Sec:SWIPT_PP_Modulation_Design} to transmissions with longer lengths for the channel input $\pmb{x}^n$. In designing coded modulation for SWIPT transmissions, we follow a similar approach presented in Section \ref{Sec:SWIPT_PP_Modulation_Design}. We first study the results obtained using learning tools. Later, we utilize the results obtained via the learning approach to design algorithmic coded modulation for PP-SWIPT.

\vspace{-1mm}
\subsection{Learning Coded Modulation for PP-SWIPT}\label{Sec:Coded_Design_Learning}
For a message set of size $M$ with channel input length $n$, the communication rate is $\tau = \frac{k}{n}$, where $k\triangleq \log{M}$. In order to make the performance results (delivered power $\bar{P}_d$ versus complementarity SER) comparable, we train the system for different values of $n$, while keeping the rate $\tau$ fixed. Here, we have trained the system for the transmission rate $\tau=2$. Given the transmission rate, the system is trained for different values of $n$ (and accordingly $k=\tau n$), namely $n=1,2,3$. Note that the message set $M=2^k=2^{\tau n}$ increases exponentially by $n$, and therefore, due to the curse of dimensionality, training the system for larger values of $n$ gets extremely demanding.

\begin{figure}[!tbp]
  \centering
  \begin{minipage}[b]{0.49\textwidth}
    \includegraphics[width=\textwidth]{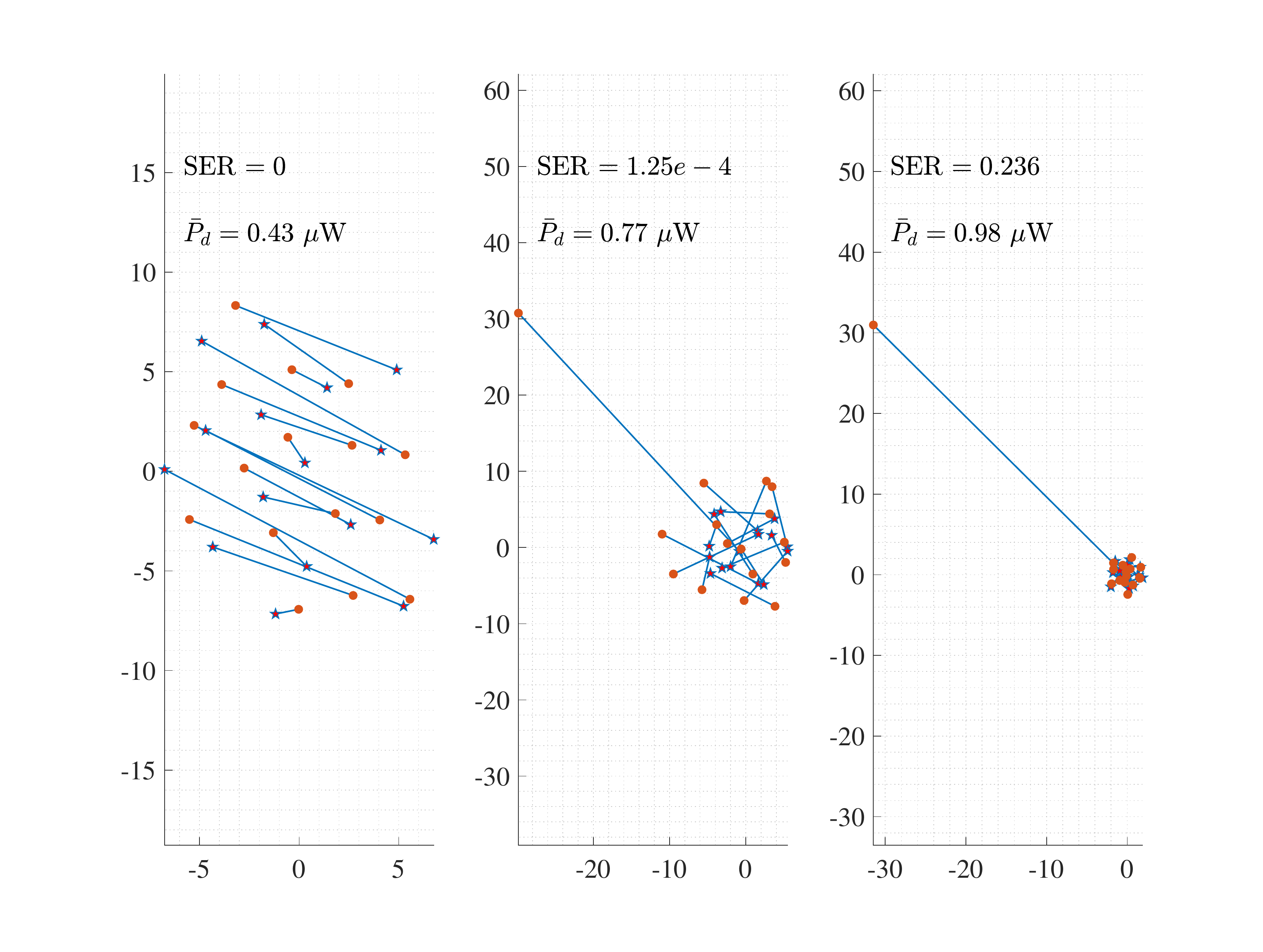}\vspace{-3mm}
    \caption{Vector representation of the learned code ($n=2$, $k=4$) with $P_a=5~\mu$W. Dots and stars represent the first and second transmitted symbols, respectively.}\label{Fig_Vector_representation}
  \end{minipage}
  \hfill
  \begin{minipage}[b]{0.49\textwidth}
    \includegraphics[width=\textwidth]{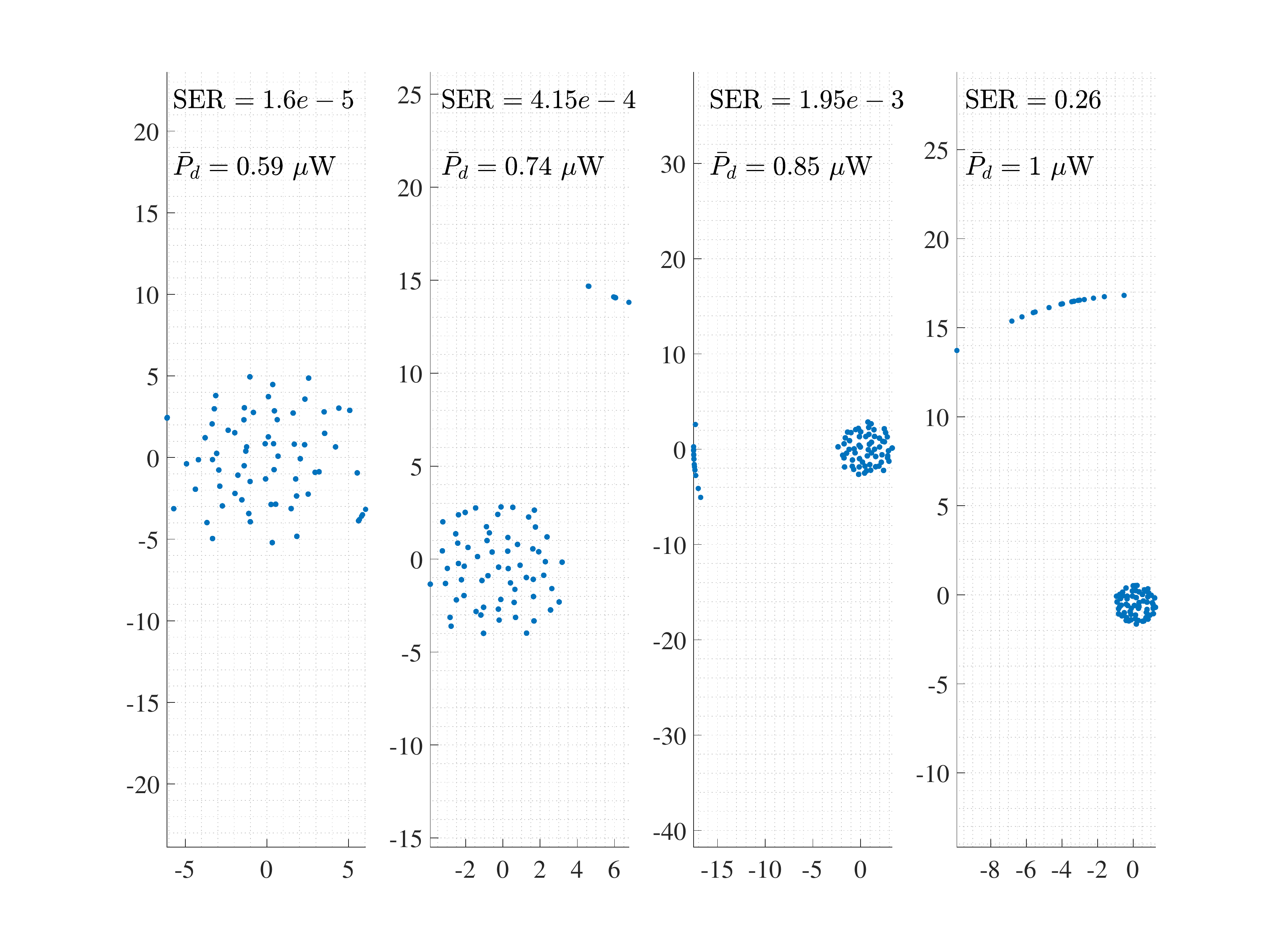}\vspace{-3mm}
    \caption{MDS representation of the learned codebook ($n=3$, $k=6$) with $P_a=5~\mu$W. Each dot represent a codeword (consisting of $3$ complex symbols) of the learned codebook.}\label{Fig_MDS_representation}
  \end{minipage}
  \vspace{-6mm}
\end{figure}

In Figure \ref{Fig_Vector_representation}, the vector representation of the trained codebook for $n=2,~k=4$ ($\tau=2$)  is illustrated for different values of power demand at the receiver. Each blue line represents a codeword. A dot and a star on each blue line represent the first and the second complex symbols transmitted for a message. It is observed that similar to the modulation design scenario, as the power demand at the receiver increases, some of the symbols of the codebook (here only one) gets further from the origin forcing the other symbols of the codebook to converge towards zero in order to satisfy the average power constraint. Similarly, in Figure \ref{Fig_MDS_representation}, the Multi-Dimensional Scaling (MDS) representation of the codewords for $n=3,~k=6$ ($r=2$) is illustrated for different power demands at the receiver (different values of $\lambda$). Note that unlike Figure \ref{Fig_Vector_representation}, here, each blue dot represents a codeword of length $n=3$ complex symbols for a message of length $k=6$ bits.

%\begin{figure}
%\begin{centering}
%\includegraphics[scale=0.3]{Vector_representation.pdf}
%\caption{Vector representation of the learned code for $n=2$ and $k=4$ under an average power constraint of $P_a=5~\mu$W. The dots and the stars represent the complex symbols transmitted as the first and second symbol of each codeword, respectively.}\label{Fig_Vector_representation}
%\par\end{centering}
%\vspace{0mm}
%\end{figure}
%
%\begin{figure}
%\begin{centering}
%\includegraphics[scale=0.3]{MDS_representation.pdf}
%\caption{MDS representation of the learned codebook for $n=3$ and $k=6$ under an average power constraint of $P_a=5~\mu$W.}\label{Fig_MDS_representation}
%\par\end{centering}
%\vspace{0mm}
%\end{figure}

\vspace{-1mm}
\subsection{Algorithmic Coded Modulation for PP-SWIPT}\label{Sec:Coded_Design_Algorithmic}
If the receiver continuously changes its information-power demand, training the system to obtain the corresponding codebook could be a time demanding and energy consuming process. This issue gets even more crucial if the focus of the application is on low-power wireless devices, where the energy consumption is critical. Accordingly, the use of learning approach for real time applications yet seems a challenge. However, the learning results can still be useful in two perspectives. First, the learned results can be used in order to come up with a heuristic algorithm that could in turn save the computation time and energy required for training. Second, the performance obtained via the learning approach can be considered as a baseline to compare the performance of a designed algorithm.

Here, inspired by the results obtained via the learning approach, we propose an algorithmic approach towards coded modulation for PP-SWIPT that achieves the aforementioned two goals satisfactorily, that is, the algorithm is easily adapted to new information-power demands of the receiver (without the need for training) and performs very close to the performance obtained by learning.

The proposed algorithm has two phases, namely, a) obtaining information codebook, b) constructing the SWIPT codebook from information codebook. In the first phase a codebook is obtained by merely focusing on the information transmission perspective. Once the information codebook is obtained, in the second phase, depending on the average power constraint at the transmitter and the power demand at the receiver, a number of codebook symbols are moved away from the origin (equivalently, forcing the other symbols of the codebook converging towards zero amplitude). The details of the algorithm are as follow.

\begin{itemize}
\item \textit{Information codebook}:
\begin{enumerate}
\item Given the message set size $M$, the length of the codewords $n$ and the average power constraint $P_a$, a constellation of $Mn$ points is created using the approach presented in \ref{Sec:Information_Only_Modulation}. Recall that the obtained constellation is such that the minimum distance between each point is $t$ (see equation (\ref{eq:circular})). The constellation points are indexed from $1$ to $Mn$.

\item The vector set forming the permutations of $Mn$ symbols taken $n$ at a time is produced. The first vector, denoted by $v_1$ is chosen randomly from the vector set $\mathcal{V}_1\triangleq\{v_1,\ldots, v_{\frac{Mn!}{(Mn-n)!}}\}$. Choose $d_{\text{min}}$ as the target minimum distance of any two codeword.

\item Given the minimum distance $d_{\text{min}}$ and the iteration index $i~(i\geq 1 )$, vector $v_{i+1}$ is chosen from the vector set $\mathcal{V}_{i+1}\triangleq\{\nu|\nu\in\mathcal{V}_i,|\nu-v_i|^2\geq d_{\text{min}}\}$, such that $v_{i+1} = \argmin\limits_{\nu\in\mathcal{V}_{i+1}} |v_i-\nu|^2$. The iteration is terminated if $\mathcal{V}_{I+1}=\varnothing$ for some index $I$.
    \begin{enumerate}
    \item If $I<Mn$, increase $d_{\text{min}}$ by $\epsilon$, set $i=1$ and repeat (3). The iteration is terminated if $I = Mn$.
    \item If $I>Mn$, decrease $d_{\text{min}}$ by $\epsilon$, set $i=1$ and repeat (3). The iteration is terminated if $I = Mn$.
    \end{enumerate}
\end{enumerate}

\item \textit{SWIPT codebook}:

Once the information codebook is obtained, we have $Mn$ complex symbols in the codebook. To satisfy different information-power demands of the receiver, we allow some of the symbols of the codebook to move away from the origin. Therefore, similarly to equation (\ref{Eq_10}), the number of symbols that are moved away (denoted by $M_{on}^c$) is determined as
\begin{equation}
      M_{on}^c=\argmin_{m\in\{1,\ldots,Mn\}} \left|p_{on}^{\star}-\frac{m}{Mn}\right|.
\end{equation}

\begin{figure}
\begin{centering}
\includegraphics[scale=0.4]{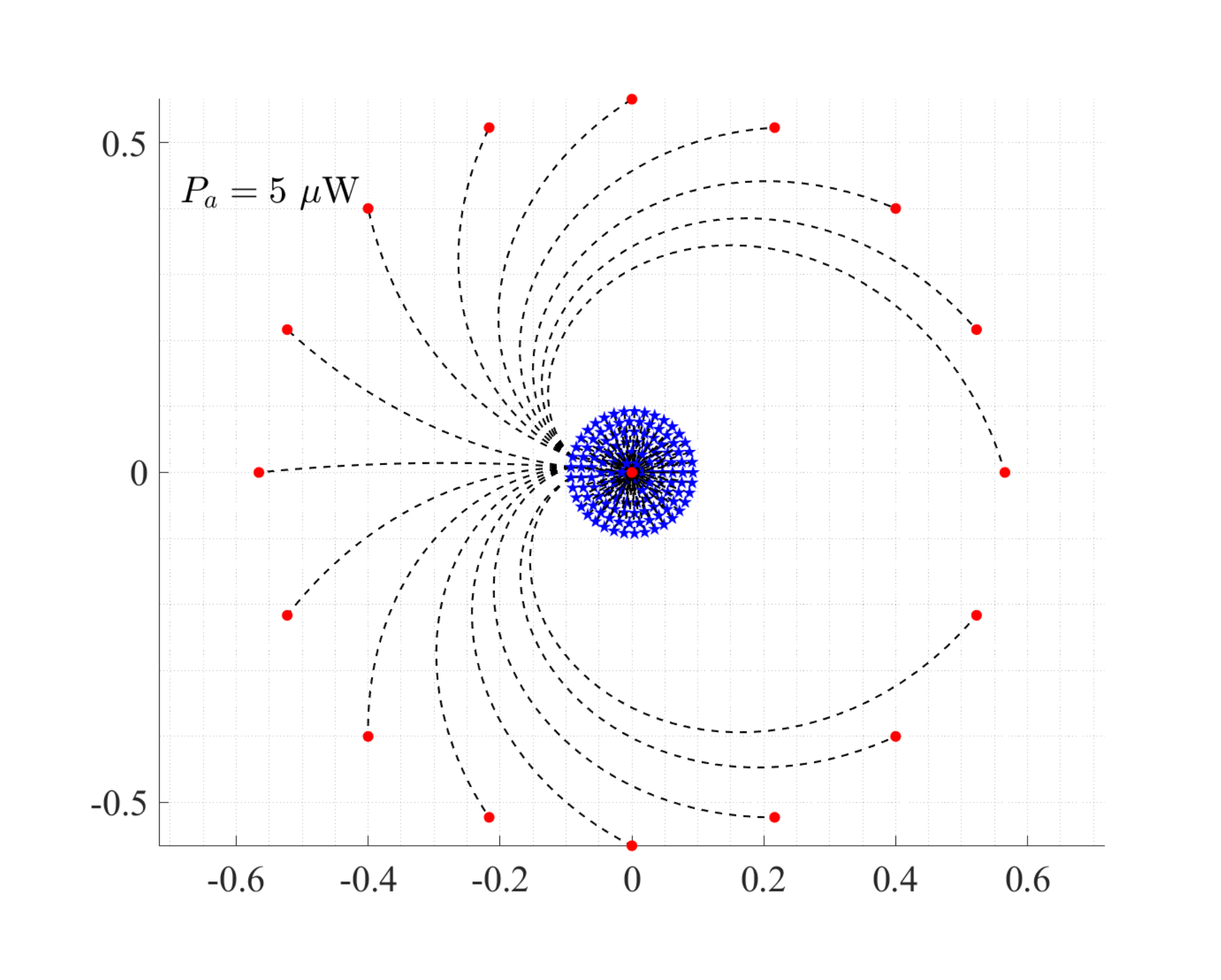}\vspace{-3mm}
\caption{Evolution of designed coded modulation from maximum information demand (blue stars) to maximum power demand (red dots) for $(n,k)=(4,8)$ and $P_a=5~\mu$W. The number of symbols that move away from zero depend on the average power constraint.}\label{Fig_n_length_modulation_evolution}
\par\end{centering}
\vspace{-5mm}
\end{figure}

As explained in Section \ref{Sec:SWIPT_Modulation}, in the extreme scenario, where the power demand is at its maximum, the transmitter can resemble the optimal On-Off keying signalling (with the probability of the ON signal equal to $p_{on}^{\star}$) by simply mapping all the $M_{on}^c$ complex symbols to the amplitude $|\pmb{x}|=\sqrt{MnP_a/M_{on}^c}$ (placed equidistantly) and the remaining points to zero amplitude. The $M_{on}^c$ complex points are chosen and moved away exactly in the same manner as that explained in Section \ref{Sec:SWIPT_Modulation}. Accordingly, for brevity, we neglect the details.

In Figure \ref{Fig_n_length_modulation_evolution}, an evolution of a designed SWIPT codebook is illustrated for $(n,k)=(4,8)$ under the average power constraint $P_a=5~\mu$W. The blue stars represent the symbols of the codebook for information demand of the receiver only. As the receiver power demand increases, some symbols of the codebook (red dots) start moving away from the origin in a way that ultimately they get around a circle with equal distance with respect to their corresponding adjacent symbols. Equivalently, the other symbols of the codebook move towards zero amplitude in order to satisfy the average power constraint of the transmitter.

\end{itemize}

\begin{figure}[!tbp]
  \centering
  \begin{minipage}[b]{0.49\textwidth}
    \includegraphics[width=\textwidth]{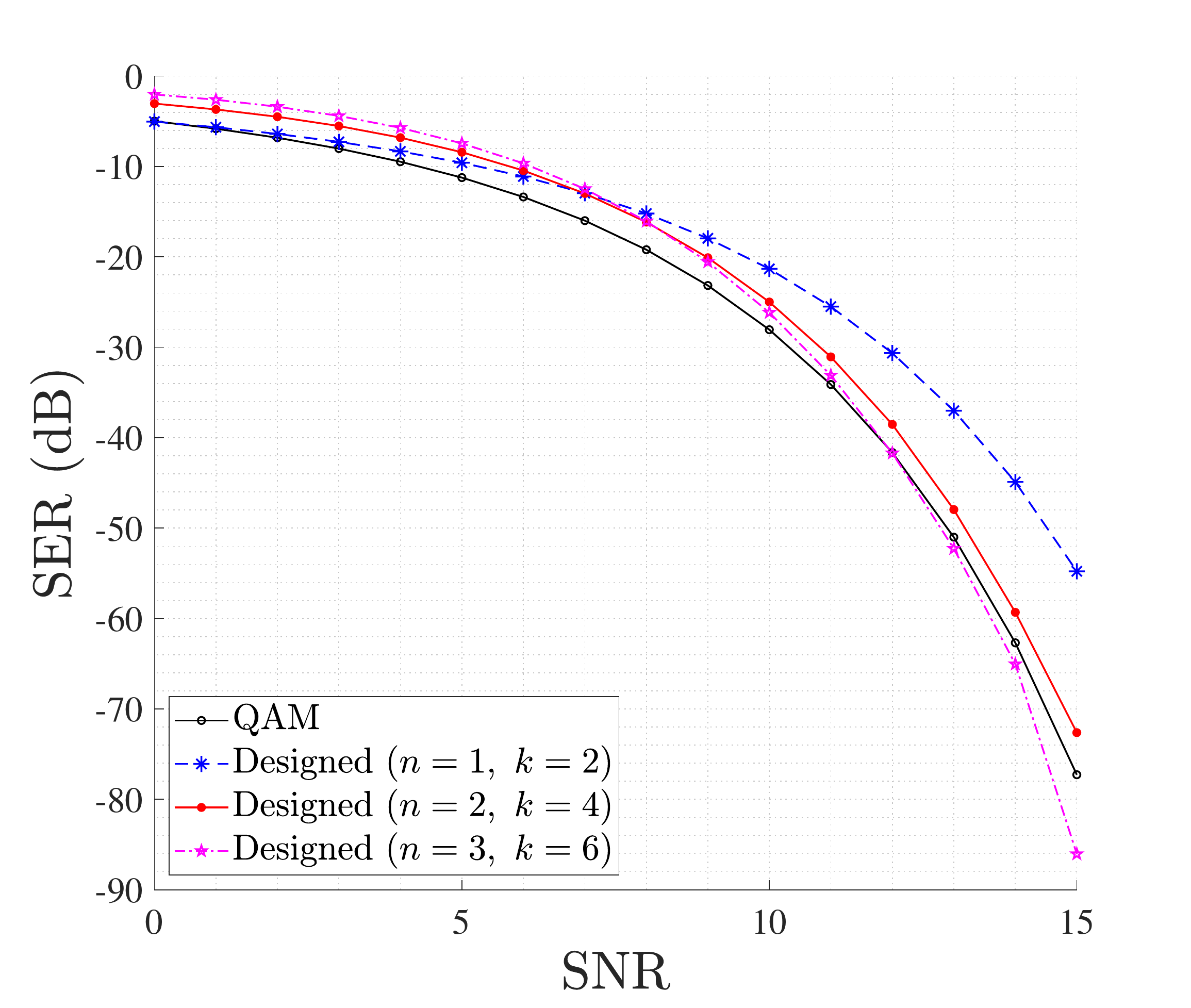}\vspace{-3mm}
    \caption{Comparison of SER vs. SNR for conventional 4-QAM, and the designed coded modulations for $(n,k)=(1,2),~(2,4)$ and $(3,6)$. The rate of the information for all the plots is $\tau=k/n=2$.}\label{Fig_SER_sv_SNR_rate_1}
  \end{minipage}
  \hfill
  \begin{minipage}[b]{0.49\textwidth}
    \includegraphics[width=\textwidth]{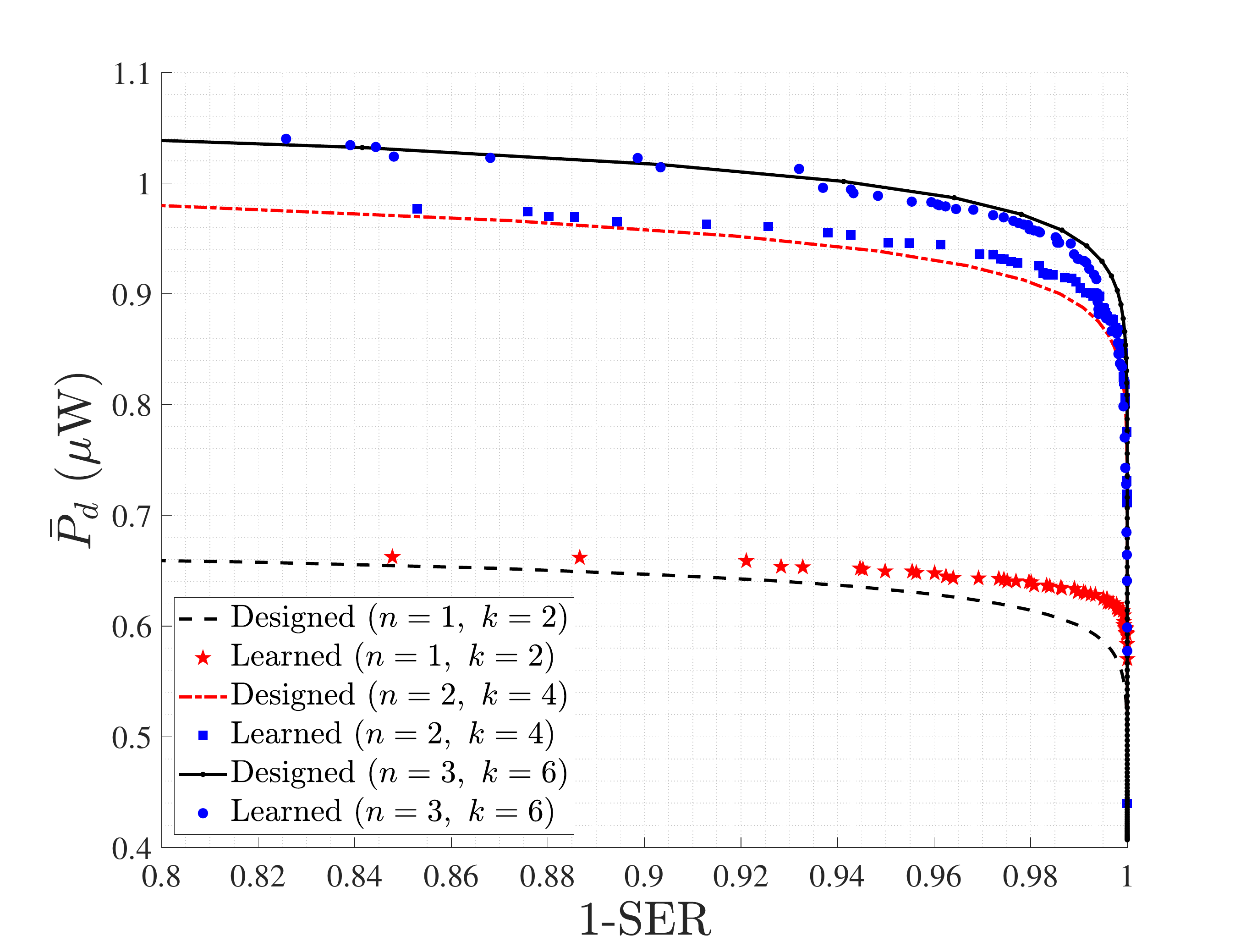}\vspace{-3mm}
    \caption{Delivered power $\bar{P}_d$ versus complementary SER ($1-\text{SER}$) tradeoff for $(n,k)=(1,2),~(2,4)$ and $(3,6)$ with $P_a=5~\mu$W. Dots and lines represent the learning and algorithmic performances, respectively.}\label{Fig_n_length_modulation_performance_rate_2}
  \end{minipage}
  \vspace{-5mm}
\end{figure}

\subsection{Performance}\label{Sec:Coded_Design_Performance}
In Figure \ref{Fig_SER_sv_SNR_rate_1}, the SER versus SNR of the different coded modulations (for $(n,k)=(1,2),~(2,4)$ and $(3,6)$) obtained via algorithmic approach are illustrated and compared with the conventional QAM modulation, i.e.,  $(n,k)=(1,2)$. The rate of all the transmissions are the same, i.e., $\tau=k/n=2$. It is observed that the information coded modulations perform close to QAM modulation with a slight improvement for $n=3$ in the high SNR regime.

In the following, the performance of the designed SWIPT coded modulation (obtained via learning and algorithmic approach) is compared. In Figure \ref{Fig_n_length_modulation_performance_rate_2}, the delivered power versus complementary SER (the dots and the lines corresponds to learning and algorithmic approach, respectively) is illustrated for for $(n,k)=(1,2),~(2,4)$ and $(3,6)$. It is observed that the region ($\bar{P}_d$ versus $1-$SER) is enlarged\footnote{Note that from SWIPT perspective, the ultimate goal (for any information-power demand) is to decrease the SER and increase $\bar{P}_d$.} by the length of the codewords $\pmb{x}^n$. This highlights the fact that coding can improve information transmission while keeping the delivered power level high. The codebooks obtained via the algorithmic approach perform very close to the codebooks obtained via the learning approach. Interestingly, for higher codeword lengths the algorithmic codes perform even better than the learned codes. This is justified by noting that for larger codeword lengths, higher complexity (in terms of the number of layers and activation functions) for the deployed NN structures is needed. Note that for a fixed transmission rate, the number of messages increases exponentially with the length of the codeword, and accordingly, for large $n$, NN structures are doomed by the curse of dimensionality. While learning codebooks for large $n$ is not feasible, the algorithmic coded modulation design is significantly fast. As a by-product, the codebook is adaptive to the fast change of the system design parameters (change of the receiver information-power demands).

One important observation in Figure \ref{Fig_n_length_modulation_performance_rate_2} is that the improvement in the region ($\bar{P}_d$ versus $1-$SER) by changing the codeword length from $n=1$ to $n=2$ compared to changing the codeword length from $n=2$ to $n=3$ is significant. Accordingly, the improvement gain is negligible for larger values of $n$.

%\begin{figure}
%\begin{centering}
%\includegraphics[scale=0.3]{SER_sv_SNR_rate_2.pdf}
%\caption{Comparison of SER vs. SNR for conventional 4-QAM, with ($(n,k)=(1,2)$), and the designed coded modulations for $(n,k)=(1,2),~(2,4)$ and $(3,6)$. The rate of the information for all the plots is $\tau=k/n=2$.}\label{Fig_SER_sv_SNR_rate_1}
%\par\end{centering}
%\vspace{0mm}
%\end{figure}
%
%
%
%\begin{figure}
%\begin{centering}
%\includegraphics[scale=0.3]{n_length_modulation_performance_rate_2.pdf}
%\caption{Illustration of the tradeoff between the average delivered power $\bar{P}_d$ and complementary symbol error rate $1-\text{SER}$ for $(n,k)=(1,2),~(2,4)$ and $(3,6)$ under an average power constraint $P_a=5~\mu$W. Black (squares, stars and circles) dots and blue (solid and dashed) lines represent the performance corresponding to learning and algorithmic approaches, respectively. The rate for all the transmissions is $\tau=k/n=2$. The region is enlarged by the length of the transmitted codeword $\pmb{x}^n$.}\label{Fig_n_length_modulation_performance_rate_2}
%\par\end{centering}
%\vspace{0mm}
%\end{figure}

\section{Conclusion}\label{Sec:Conclusion}
In this paper, we have studied SWIPT signal design accounting for the practical nonlinearity of the EH over the entire range of its input power. We first obtained a parametric EH model by applying nonlinear regression over real data collected from a practical EH. Unlike the proposed models in the literature, the learned model captures the practical limitations of an EH in a unified manner. Considering the obtained EH model, we studied learning modulation design for PP-SWIPT. Inspired by the learning results, an algorithmic approach for modulation design is proposed, which performs close to the learning approach. The algorithmic approach does not require training, and accordingly, is highly adaptive to the fast change of system design parameters. We extend the learning approach for modulation design for multi-user SWIPT and coded modulation design for PP-SWIPT. In particular, for PP-SWIPT coded modulation, inspired by the learning results, we propose an algorithm that performance-wise is almost the same as the learning approach with faster adaptivity against the change of system design parameters.

The use of ML to better understand designing communications systems is rather a new topic, and potentially can be extended to many interesting problems. This can be considered in two directions. First, to get intuitions from the learning results to design heuristic algorithms, which bypasses the training load of the learning approach while achieving approximately the same performance. Second, in the absence of tight information theoretic results for fundamental performance limits of short block transmissions, complex deep learning algorithms can be considered as an achievable milestone in designing communications systems.

%\bibliographystyle{IEEEtran}
%\bibliography{ref_1.bib}

\bibliographystyle{ieeetran}
\bibliography{ref_1}
% \bibliographystyle{ieeetran}
% \bibliography{ref}

\end{document}